\documentclass[12pt]{iopart}
\begin{document}
\title{Cosmology and Brane Worlds: A Review}
\author{Philippe Brax\dag~and Carsten van de Bruck\ddag}
\address{\dag\ Service de Physique Th\'eorique, CEA-Saclay
F-91191, Gif/Yvette cedex, France}
\address{\ddag\ Department of Applied Mathematics and Theoretical Physics,\\ 
~~Center for Mathematical Sciences, University of Cambridge, \\
~~Wilberforce Road, Cambridge CB3 0WA, U.K.\\ 
~~and \\
~~Astrophysics Group, University of Oxford\\
~~Denys Wilkinson Building\\
~~Keble Road\\
~~Oxford, OX1 3RG, U.K.}
\ead{brax@spht.saclay.cea.fr, cvdb@astro.ox.ac.uk}
\begin{abstract}
Cosmological consequences of the brane world 
scenario are reviewed in a pedagogical manner. According to 
the brane world idea, the standard model particles are 
confined on a hyper--surface (a so--called brane), which is embedded in a 
higher--dimensional spacetime (the so--called bulk). We begin our review with
the simplest consistent brane world model: a single brane 
embedded in a five--dimensional 
Anti-de Sitter space--time. Then we include a scalar field 
in the bulk and discuss in detail the difference with
the Anti-de Sitter case.  The geometry of 
the bulk space--time is also analysed in some depth. Finally, we investigate  
the cosmology of a system with two branes and a bulk scalar 
field. We comment on brane collisions and summarize some open 
problems of brane world cosmology. 
\end{abstract}

\maketitle

\section{Introduction}
The idea of extra dimensions was proposed in the 
early twentieth century by 
Nordstrom and a few years later by Kaluza and Klein \cite{Kaluza}. 
It has reemerged over the years  in theories  combining  the 
principles of quantum mechanics and relativity. In particular
theories based on supersymmetry, especially superstring 
theories, are naturally expressed in more than four dimensions
\cite{polchinskibook}.
Four dimensional physics is retrieved by  Kaluza--Klein
reduction, i.e compactifying  on a manifold of small size,
typically much smaller than the size of an atomic nucleus. 

Recent developments in string theory and its extension M--theory
have suggested  another approach  to  compactify extra spatial 
dimensions. According to these developments, the standard model
particles are confined on a hypersurface (called brane) embedded in a 
higher dimensional space (called bulk). Only gravity and other 
exotic matter such as the dilaton 
can propagate in the bulk. 
Our universe may  be such a brane--like object. This idea was 
originally motivated phenomenologically (see \cite{akama}--\cite{gibbons})
and later revived in string theory. Within the 
brane world scenario,  constraints on the size of extra dimensions become weaker, 
because the standard model particles propagate only in three spatial dimensions.
Newton's law of gravity, however, is sensitive to the presence of
extra--dimensions. Gravity is being tested only on scales larger than a tenth 
of a millimeter and possible deviations below that scale can be envisaged.

From the string theory point of view, 
brane worlds of the kind  discussed in this review 
spring from a model suggested by Horava and Witten \cite{horavawitten}. 
The strong coupling limit of the $E_8 \times E_8$
heterotic string theory at low energy is described by 
eleven dimensional supergravity with the eleventh dimension
compactified on an orbifold with $Z_2$ symmetry, i.e. an interval. The
two boundaries of  
spacetime (i.e.  the orbifold fixed points) are 10--dimensional planes, 
on which gauge theories (with the $E_8$ gauge groups) are
confined. Later Witten argued that 6 of the 11 dimensions can 
be consistently compactified on a Calabi--Yau threefold and that the 
size of the Calabi-Yau manifold can be substantially smaller than the space
between the two boundary branes \cite{witten}. Thus, in that limit space--time
looks five--dimensonal with four dimensional boundary 
branes \cite{lukasstelle}. This provides the underlying 
picture for many brane world models proposed so far. 

Another important ingredient  was put forward by  Arkani-Hamed, 
Dimopoulos and Dvali (ADD), \cite{arkanihamed1} and \cite{arkanihamed2},
following an earlier idea by Antoniadis \cite{antoniadis}, 
who suggested that by confining  the standard model 
particle  on a brane the extra dimensions can be larger 
than previously anticipated. They considered a  
flat bulk geometry in  ($4+d$)--dimensions, in which $d$ dimensions 
are compact with radius $R$ (toroidal topology). The 
four--dimensional Planck mass $M_P$ and the ($4+d$)--dimensional 
Planck mass $M_{\rm fund}$, the gravitational scale of the extra 
dimensional theory, are related by
\begin{equation}
M_{P}^2 = M_{\rm fund}^{2+d} R^d.
\end{equation}
Gravity  deviates from Newton's law only on scales smaller than $R$. 
Since gravity is tested only down to sizes of around a
millimeter, $R$ could be as large as a fraction of a millimeter. 

ADD assumed that the bulk geometry is flat. Considerable progress
was made by Randall and Sundrum, who considered {\it non--flat}, 
i.e. warped bulk geometries \cite{randallsundrum1}, \cite{randallsundrum2}. 
In their models, the bulk spacetime 
is a slice of Anti--de Sitter spacetime, i.e. a space-time with a
negative cosmological constant. Their discovery was that, due to 
the curvature of the bulk space time, Newton's law of gravity can 
be obtained on the brane of positive tension embedded in an 
infinite extra--dimension. Small corrections to Newton's law
are generated and constrain the possible scales in the model to
be smaller than a millimetre. 

They also proposed a two--brane model 
in which the hierarchy problem, i.e. the large discrepancy between the
Planck scale at $10^{19}$ GeV and the electroweak scale at $100$ GeV, 
can be addressed. The large hierarchy is due to the
highly curved AdS background which implies a large gravitational red-shift
between energy scale on the two branes. In this scenario, the standard 
model particles are confined on a brane with negative tension sitting 
at $y=r_c$, whereas a positive tension brane is located at $y=0$. 
The large hierarchy is generated by the appropriate inter-brane
distance, i.e. the  radion. It can be shown that the 
Planck mass $M_{\rm Pl}$ measured on the negative tension brane is given by 
($k = \sqrt{-\Lambda_5\kappa_5^2/6}$),
\begin{equation}
M_{\rm Pl}^2 \approx e^{2kr_c}M_5^3/k,
\end{equation}
where $M_5$ is the five--dimensional Planck mass and $\Lambda_5$ the 
(negative) cosmological constant in the bulk. Thus, we see that, if
$M_5$ is not very far from the electroweak scale $M_W \approx $TeV, we need 
$kr_c \approx 50$, in order to generate a large Planck mass on our brane.
Hence, by tuning the radius $r_c$ of the extra dimension to a reasonable
value, one can obtain a very large hierarchy between the weak and
the Planck scale. Of course, a complete
realization of this mechanism requires an explanation for such a
value of  the radion. In other words, the radion needs to be
stabilized at a certain value. The stabilization mechanism is not 
thoroughly understood, though models with a bulk scalar field have 
been proposed and have the required properties \cite{goldbergerwise}. 

Another puzzle which might be addressed with brane models is the cosmological
constant problem. One may invoke an extra dimensional origin for  the apparent
(almost) vanishing of the cosmological constant. The self-tuning
idea \cite{selftuning} advocates that the energy density on our brane does
not lead
to a large curvature of our universe. On the contrary, the extra
dimension becomes highly curved, preserving a flat Minkowski brane with
apparent vanishing cosmological constant. Unfortunately, the
simplest realization of this mechanism with a bulk scalar field
fails due to the presence of a naked singularity in the bulk.
This singularity can be shielded by a second brane whose tension
has to be fine-tuned with the original brane tension \cite{nilles}. In a sense,
the fine tuning problem of the cosmological constant reappears through
the extra dimensional back-door.

Finally, we will later discuss in some detail another 
spectacular consequence of brane cosmology, namely the
possible modification to the Friedmann equation at very high 
energy \cite{binetruy1}. This effect was first recognised in 
\cite{boundaryinflation} in the context of inflatonary solutions. 
As we will see, Friedmann's equation has, for the Randall--Sundrum 
model, the form (\cite{grojean} and \cite{csaki}) 
\begin{equation}
H^2=\frac{\kappa_5^4}{36}\rho^2 + \frac{8\pi G_N}{3}\rho + \Lambda,
\end{equation}
relating the expansion rate of the brane $H$ to the (brane) matter 
density $\rho$ and the (effective) cosmological constant $\Lambda$. 
The cosmological constant can be tuned to zero by an appropriate choice of
the brane tension and bulk cosmological constant, as in the Randall-Sundrum 
case. Notice that at high energies, for which 
\begin{equation}
\rho \gg \frac{96 \pi G_N}{\kappa_5^4},
\end{equation}
where $\kappa_5^2$ is the five dimensional gravitational constant,
the Hubble rate becomes
\begin{equation}
H \propto \rho,
\end{equation}
while in ordinary cosmology $H\propto \sqrt \rho$. The latter
case is retrieved at low energy, i.e.
\begin{equation}
\rho \ll \frac{96 \pi G_N}{\kappa_5^4},
\end{equation}
Of course modifications to the Hubble rate can only be
significant before nucleosynthesis. They may have drastic consequences
on early universe phenomena such as inflation.

In this article we will review these and other aspects of the 
brane world idea in the context of cosmology. In order to give a 
pedagogical introduction to the subject, we will follow a 
phenomenological approach and start with the simplest model, 
i.e. the Randall-Sundrum model, with a brane embedded in a 
five--dimensional vacuum bulk spacetime (section 2). Later 
we will include a bulk scalar field (section 3). In section 4 
we will discuss the geometry of the bulk--spacetime in some 
detail. In the last part we will discuss more realistic models 
with two branes and bulk scalar fields (section 5). In section 
6 we will discuss brane collisions. Open questions are summarized in 
section 7. We would like to mention other review articles on brane 
worlds and cosmology, taking different approaches from the one 
taken here \cite{luki}-\cite{quevedo}. We will mostly be concerned 
with the case, in which the bulk space--time is five--dimensional.

\section{The Randall--Sundrum Brane World}
Originally, Randall and Sundrum suggested a two--brane scenario 
in five dimensions with a highly curved  bulk geometry as an explanation for
the large hierarchy between the  Planck scale 
and the electroweak energy--scale \cite{randallsundrum1}. 
In this scenario, the standard 
model particles live on a brane with (constant) negative tension, 
whereas the bulk is a slice of  Anti--de Sitter (AdS) spacetime , i.e. 
a space-time with a negative cosmological constant. In the bulk 
there is another brane with positive tension. This is the so--called 
Randall--Sundrum I (RSI) model.  
Analysing the solution of Einstein's
equation on the  positive tension brane
and sending the negative tension brane to infinity, an observer 
confined to the positive tension brane recovers Newton's 
law if the curvature scale of the AdS is smaller than a
millimeter \cite{randallsundrum2}. The higher--dimensional space is  
{\it non--compact}, which must be contrasted with the Kaluza--Klein
mechanism, where all extra--dimensional degrees of freedom are
compact. This one--brane model, on which we will concentrate in this 
section, is the so--called Randall--Sundrum II (RSII) model. 
It was shown, there is a {\it continuum} of Kaluza--Klein
modes for the gravitational field, contrasting with the discrete spectrum 
if the extra dimension is  periodic.
This leads to a correction to the force between two static masses
on the brane. To be specific, it was shown that the potential energy
between two point masses confined on the brane is given by
\begin{equation}
V(r) = \frac{G_N m_1 m_2}{r}\left(1 + \frac{l^2}{r^2} + O(r^{-3})\right).
\end{equation}
In this equation, $l$ is related to the five--dimensional bulk cosmological 
constant $\Lambda_5$ by $l^2 = -6/(\kappa_5^2\Lambda_5)$ 
and is therefore a measure of the
curvature scale of the bulk spacetime. Gravitational experiments 
show no deviation from Newton's law of gravity on length scales 
larger than a millimeter \cite{submillimeter}. Thus, $l$ has to be 
smaller than that length scale.

The static solution of the Randall and Sundrum model  
can be obtained as follows: The total action
consists of the Einstein-Hilbert action and the 
brane action, which in the Randall--Sundrum model 
have the form
\begin{eqnarray}
S_{\rm EH} &=& -\int dx^5 \sqrt{-g^{(5)}} 
\left( \frac{R}{2\kappa_5^2} + \Lambda_{5} \right), \\
S_{\rm brane} &=& \int dx^4 \sqrt{-g^{(4)}}\left( -\sigma \right). 
\end{eqnarray}
The parameter $\Lambda_5$ (the bulk cosmological constant) 
and $\sigma$ (the brane tension) are constant and $\kappa_5$ 
is the five--dimensional gravitational coupling constant. The 
brane is located at $y=0$ and we assume a $Z_2$ symmetry, i.e. 
we identify $y$ with $-y$. The ansatz for the metric is
\begin{equation}
ds^2 = e^{-2 K(y)}\eta_{\mu\nu} dx^{\mu}dx^{\nu} + dy^2.
\end{equation}
Einstein's equations, derived from the action above, give two 
independent equations:
\begin{eqnarray}
6K'^2 &=& - \kappa_5^2 \Lambda_{5} \nonumber \\
3K'' &=& \kappa_5^2 \sigma \delta(y) \nonumber.
\end{eqnarray}
The first equation can be easily solved:
\begin{equation}\label{K}
K = K(y) = \sqrt{-\frac{\kappa_5^2}{6}\Lambda_5}\ y \equiv ky, 
\end{equation}
which tells us that $\Lambda_5$ must be negative. If we integrate 
the second equation from $-\epsilon$ to $+\epsilon$, take the limit 
$\epsilon \rightarrow 0$ and make use of the $Z_2$--symmetry, we get 
\begin{equation}
6K'\vert_0 = \kappa_5^2 \sigma
\end{equation}
Together with eq. (\ref{K}) this tells us that 
\begin{equation}\label{finetuning}
\Lambda_5 = -\frac{\kappa_5^2}{6}\sigma^2
\end{equation}
Thus, there must be a fine--tuning between the brane 
tension and the bulk cosmological constant  for 
static solutions to exist.  In this section we will discuss the
cosmology of this model in detail. 

\subsection{Einstein's equations on the brane}
There are two ways of  deriving  the cosmological equations and 
we will describe both of them below. The first one is rather 
simple and makes use of the bulk equations only. The second 
method uses the  geometrical relationship between four--dimensoinal 
and five--dimensional quantities. 
We begin with the simpler method.

\subsubsection{ Friedmann's equation from 
five--dimensional Einstein equations}
In the following subsection we will set $\kappa_5 \equiv 1$.
We write the bulk metric as follows:
\begin{equation}\label{metricbraneframe}
ds^2 = a^2 b^2(dt^2 - dy^2) - a^2 \delta_{ij}dx^i dx^j.
\end{equation}
This metric is consistent with homogeneity and isotropy 
on the brane located at $y=0$. The functions $a$ and $b$ are 
functions of $t$ and $y$ only. Furthermore, we have assumed 
flat spatial sections, it is straightforward to include a 
spatial curvature. For this metric, Einstein equations 
in the bulk read:
\begin{eqnarray}
a^2b^2{G^0}_0 &\equiv& 3\left[2\frac{\dot{a}^2}{a^2}+\frac{\dot{a}\dot{b}}{ab}
                   -\frac{a''}{a}+\frac{a' b'}{ab}+kb^2\right]
                = a^2b^2\left[\rho_B +\rho^{ }
                \bar\delta(y-y_b)\right]\label{G00}\\
a^2b^2{G^5}_5 &\equiv&3\left[\frac{\ddot{a}}{a}-\frac{\dot{a}\dot{b}}{ab}
                 -2\frac{{a'}^2}{a^2}-\frac{a' b'}{ab}+kb^2\right]
                 =-a^2b^2T^5_5 \label{G55}\\
a^2b^2{G^0}_5&\equiv&3\left[ -\frac{\dot{a}'}{a}+2\frac{\dot{a}a'}{a^2}
                 +\frac{\dot{a}b'}{ab}+\frac{a'\dot{b}}{ab}\right]
                 = -a^2b^2 T^0_5  \label{G05}\\
a^2b^2{G^i}_j&\equiv&\left[ 3\frac{\ddot{a}}{a}+\frac{\ddot{b}}{b}-
                 \frac{\dot{b}^2}{b^2}-3\frac{a''}{a}-\frac{b''}{b}
                 +\frac{{b'}^2}{b^2}+kb^2\right]
                 {\delta^i}_j \nonumber \\
                 &=&-a^2b^2\left[p_B+p^{ }
                 \bar{\delta}(y-y_b)\right]{\delta^i}_j ,\label{Gij}
\end{eqnarray}
where the bulk
energy--momentum tensor $T^a_b$ has been kept general here. For the Randall--Sundrum model 
we will now take $\rho_B = - p_B = \Lambda_5$ and $T^0_5=0$. 
Later, in the next section, we will use these 
equations to derive  Friedmann's equation with 
a bulk scalar field. In the equations above, a dot represents 
the derivative with respect to $t$ and a prime a derivative with 
respect to $y$. 

Let us integrate the 00--component over $y$ from $-\epsilon$ to 
$\epsilon$ and use the fact that $a(y)=a(-y)$, $b(y)=b(-y)$,
$a'(y)=-a(-y)$ and $b'(y)=-b(-y)$ (i.e. $Z_2$-symmetry). Then, taking 
the limit $\epsilon \rightarrow 0$ we get 
\begin{equation}\label{junc1}
\frac{a'}{a}\left|_{y=0} \right. = \frac{1}{6}ab \rho.
\end{equation}
Integrating the $ij$--component in the same way and using the last 
equation gives
\begin{equation}\label{junc2}
\frac{b'}{b}\left|_{y=0} \right. = -\frac{1}{2}ab (\rho+p).
\end{equation}
These two conditions are called the junction conditions. The other
components of the  Einstein  equations should be compatible with these 
conditions. It is not difficult to show that the restriction of 
the $05$ component to $y=0$ leads to 
\begin{equation}\label{energyRS}
\dot\rho + 3\frac{\dot a}{a}\left(\rho + p \right) = 0,
\end{equation}
where we have made use of the junction conditions (\ref{junc1}) and 
(\ref{junc2}). This is nothing but matter conservation on the
brane.

Proceeding in the same way with the 55--component gives
\begin{equation}
\frac{\ddot{a}}{a}-\frac{\dot{a}\dot{b}}{ab}+kb^2
  = -{a^2b^2\over 3}\left[\frac{1}{12}\rho^{ }\left(\rho{ }+3p^{ }\right)
    +q_B\right]\; .
\end{equation}
Changing to cosmic time $d\tau = ab dt$, writing 
$a = \exp(\alpha(t))$ and using the energy conservation gives (\cite{flanagan},
\cite{vandebruck1}) 
\begin{equation}
\frac{d(H^2 e^{4\alpha})}{d\alpha} = 
\frac{2}{3}\Lambda_5 e^{4\alpha} 
+ \frac{d}{d\alpha}\left( e^{4\alpha} \frac{\rho^2}{36}\right).
\end{equation}
In this equation $aH=da/d\tau$. This equation can easily be 
integrated to give
\begin{equation}
H^2 = \frac{\rho^2}{36} + \frac{\Lambda_5}{6} + \frac{\mu}{a^4}.
\end{equation}
The final step is to split the total energy--density and pressure into
 parts coming from matter and  brane tension, i.e. to write
$\rho = \rho_M + \sigma$ and $p = p_M - \sigma$.
Then we find  Friedmann's equation
\begin{equation}
H^2 = \frac{8\pi G}{3}\rho_M\left[1+\frac{\rho_M}{2\sigma}\right] +
\frac{\Lambda_4}{3} + \frac{\mu}{a^4}, 
\end{equation}
where we have made the identification
\begin{eqnarray}
\frac{8\pi G}{3} &=& \frac{\sigma}{18} \\
\frac{\Lambda_4}{3} &=& \frac{\sigma^2}{36} + \frac{\Lambda_5}{6}.
\end{eqnarray}
Comparing the last equation with the fine--tuning relation (\ref{finetuning}) 
in the static Randall--Sundrum solution, we see that $\Lambda_4 = 0$ in 
this case. If there is a small mismatch between the brane tension and 
the five--dimensional cosmological constant, then an effective 
four--dimensional cosmological constant is generated. Another 
important point is that the four--dimensional Newton constant 
is directly related to the brane tension. The constant $\mu$
appears in the derivation above as an integration constant. 
The term including $\mu$ is called the {\it dark radiation} term 
(see e.g. \cite{darkradiation1}-\cite{darkradiation3}).
The parameter $\mu$ can be obtained from a full analysis of the 
bulk equations \cite{shirubulk}-\cite{binetruy2} (we will discuss this
in section 4). 
An extended version of Birkhoff's theorem tells us
that if the bulk spacetime is AdS, this constant is zero 
\cite{charmousis}. If 
the bulk is AdS--Schwarzschild instead, $\mu$ is non--zero but a 
measure of the mass of the bulk black hole. In the following we will assume 
that $\mu=0$ and $\Lambda_4=0$. 

The most important change in Friedmann's equation compared to the 
usual four--dimensional form is the appearance of a term proportional
to $\rho^2$. It tells us 
that if the matter energy density is much larger than the brane
tension, i.e. $\rho_M \gg \sigma$, the expansion rate $H$ is 
proportional $\rho_M$, instead of $\sqrt{\rho_M}$. The expansion rate 
is, in this regime, larger in this brane world scenario. Only in the 
limit where the brane tension is much larger than the matter energy
density, the usual behaviour $H \propto \sqrt{\rho_M}$ is
recovered. This is the most important change in brane world
scenarios. It is quite generic and not restricted to the 
Randall--Sundrum brane world model. From Friedmann's
equation and from the energy--conservation equation we can derive 
Raychaudhuri's equation:
\begin{equation}
\frac{dH}{d\tau} = -4\pi G (\rho_M + p_M)\left[ 1+ \frac{\rho_M}{\sigma}
\right].
\end{equation}
We will use these equations later in order to investigate  
inflation driven by a scalar field confined on the brane.

Notice that 
at the time of nucleosythesis the brane world corrections in Friedmann's equation 
must be negligible, otherwise the expansion rate is modified and 
the results for the abundances of  the light elements are unacceptably
changed. This implies that $\sigma \ge (1 {\rm MeV})^4$. Note, however, that 
a much stronger constraint arises from current tests for deviation from Newton's law 
\cite{maartens} (assuming the Randall--Sundrum fine--tuning relation (\ref{finetuning})): 
$\kappa_5^{-3} > 10^5$ TeV and $\sigma \ge (100 {\rm GeV})^4$. Similarily, 
cosmology constrains the amount of dark radiation. It has been shown that 
the energy density in dark radiation can at most be 10 percent of the 
energy density in photons \cite{darkradiation2}.

\subsubsection{Another derivation of Einstein's equation}
There is a more powerful way of  deriving Einstein's equation on 
the brane \cite{shiromizu}. Consider an arbitrary (3+1) dimensional 
hypersurface ${\cal M}$ with unit normal vector $n_a$ embedded 
in a 5 dimensional spacetime. The induced metric and the 
extrinsic curvature of the hypersurface are 
defined as 
\begin{eqnarray}
h^a_{~b} &=& \delta^a_{~b} - n^a n_b, \\
K_{ab} &=& h_a^{~c}h_b^{~d} \nabla_c n_d.
\end{eqnarray}
For the derivation we need three equations, two of them relate 
four--dimensional quantities constructed from $h_{ab}$ to 
full five--dimensional quantities constructed from $g_{ab}$. 
We just state these equations here and refer to \cite{wald} for 
the derivation of these equations.
The first equation is the Gauss equation, which reads
\begin{equation}
R^{(4)}_{abcd} = h_{a}^{~j}h_{b}^{~k} h_{c}^{~l}h_{d}^{~m}
R_{jklm} - 2K_{a[c}K_{d]b}.
\end{equation}
This equation relates the four--dimensional curvature tensor 
$R^{(4)}_{abcd}$, constructed from $h_{ab}$, to the five--dimensional 
one and $K_{ab}$. The next equation is the Codazzi equation, which 
relates $K_{ab}$, $n_a$ and the five--dimensional Ricci tensor:
\begin{equation}
\nabla^{(4)}_b K^{b}_{~a} - \nabla^{(4)}_a K
= n^c h^{b}_{~a}R_{bc}.
\end{equation}
One decomposes the five--dimensional curvature 
tensor $R_{abcd}$ into the Weyl--tensor $C_{abcd}$ and the 
Ricci tensor:
\begin{equation}
R_{abcd} = \frac{2}{3} \left(g_{a[c}R_{d]b} 
- g_{b[c}R_{d]a} \right) - \frac{1}{6} R g_{a[c}g_{b]d} 
+ C_{abcd}.
\end{equation}
If one substitutes  the last equation into the Gauss equation and 
constructs the four--dimensional Einstein tensor, one obtains 
\begin{eqnarray}\label{eintensor}
G^{(4)}_{ab} &=& \frac{2}{3}\left( G_{cd} h^{c}_{~a}h^{d}_{~b}
 + \left(G_{cd}n^c n^d - \frac{1}{4}G \right)h_{ab} \right) 
\nonumber \\
&+& K K_{ab} - K_a^{~c}K_{bc} - \frac{1}{2}\left(K^2 
- K^{cd}K_{cd}\right)h_{ab} - E_{ab}~,
\end{eqnarray}
where 
\begin{equation}
E_{ab} = C_{acbd}n^c n^d.
\end{equation}
We would like to emphasize that this equation holds for any 
hypersurface. If one considers a hypersurface with energy 
momentum tensor $T_{ab}$, then there exists a relationship 
between $K_{ab}$ and $T_{ab}$ ($T$ is the trace of $T_{ab}$) \cite{israel}:
\begin{equation}
\left[ K_{ab} \right] = -\kappa_5^2 \left(T_{ab} 
- \frac{1}{3} h_{ab}T \right),
\end{equation}
where $[...]$ denotes the {\it jump}: 
\begin{equation}
[f](y) = {\rm lim}_{\epsilon \rightarrow 0} 
\left( f(y+\epsilon) - f(y-\epsilon)\right).
\end{equation}
These equations are called junction conditions and are 
equivalent in the cosmological context to the junction 
conditions (\ref{junc1}) and (\ref{junc2}). Splitting 
$T_{ab} = \tau_{ab} - \sigma h_{ab}$ and inserting the junction 
condition into equation (\ref{eintensor}), we obtain
Einstein's equation on the brane:
\begin{equation}\label{einsteinbrane}
G^{(4)}_{ab} = 8\pi G \tau_{ab} - \Lambda_4 h_{ab} 
+ \kappa_5^4 \pi_{ab} - E_{ab}.
\end{equation} 
The tensor $\pi_{ab}$ is defined as 
\begin{equation}
\pi_{ab} = \frac{1}{12} \tau \tau_{ab} - \frac{1}{4} 
\tau_{ac}\tau_{b}^{~c} + \frac{1}{8} h_{ab}\tau_{cd}\tau^{cd}
- \frac{1}{24} \tau^2 h_{ab},
\end{equation}
whereas 
\begin{eqnarray}
8\pi G &=& \frac{\kappa_5^4}{6}\sigma \\
\Lambda_4 &=& \frac{\kappa_5^2}{2}\left(\Lambda_5 + 
\frac{\kappa_5^2}{6}\sigma^2\right).
\end{eqnarray}
Note that in the Randall--Sundrum case we have $\Lambda_4 = 0$
due to the fine--tuning between the brane tension and the bulk cosmological 
constant. Moreover  $E_{ab} = 0$ as  the Weyl--tensor vanishes for 
an AdS spacetime. In general, the energy conservation and the 
Bianchi identities imply that 
\begin{equation}\label{weylfluid}
\kappa_5^4 \nabla^{a} \pi_{ab} = \nabla^{a} E_{ab}
\end{equation}
on the brane.

Clearly, this method is  powerful, as it does not assume
homogeneity and isotropy nor does it assume the 
bulk to be AdS. In the case of an AdS bulk and a Friedmann--Robertson 
walker brane, the previous equations  reduce to the Friedmann equation 
and Raychaudhuri equation derived earlier. However, the set of 
equations on the brane {\it are not closed in general} \cite{roy}, 
as we will see in the next section. 

\subsection{Slow--roll inflation on the brane}
We have seen that the Friedmann equation on a brane is 
drastically modified at high energy where the $\rho^2$ terms dominate.
As a result the early universe cosmology on branes tends to be
different from standard 4d cosmology. In that vein it seems
natural to look for brane effects on early universe phenomena
such as inflation (see in particular \cite{wandsinflation} and 
\cite{copeland}) and on phase--transitions \cite{anne}. 

The energy density and the pressure of a scalar field are given by 
\begin{eqnarray}
\rho_{\phi} &=& \frac{1}{2}\phi_{,\mu}\phi^{,\mu} + V(\phi),\\
p_{\phi} &=& \frac{1}{2}\phi_{,\mu}\phi^{,\mu} - V(\phi),
\end{eqnarray}
where $V(\phi)$ is the potential energy of the scalar field. The full 
evolution of the scalar field is described by the (modified) Friedmann
equation, the Klein--Gordon equation and the Raychaudhuri equation. 

We will assume that the field is in a slow--roll regime, the evolution 
of the fields is governed by (from 
now on a dot stands for a derivative with respect to cosmic time)
\begin{eqnarray}
3H\dot\phi &\approx& - \frac{\partial V}{\partial \phi} \\
H^2 &\approx& \frac{8\pi G}{3} V(\phi)\left(1 + \frac{V(\phi)}{2\sigma}\right).
\end{eqnarray}
It is not difficult to show that these equations imply that the 
slow--roll parameter are given by
\begin{eqnarray}
\epsilon &\equiv& -\frac{\dot H}{H^2} 
= \frac{1}{16\pi G}\left(\frac{V'}{V}\right)^2\left[ 
\frac{4\sigma(\sigma + V)}{(2\sigma + V)^2} \right] \label{epsilon}\\
\eta &\equiv& \frac{V''}{3H^2}=\frac{1}{8\pi G}\left(\frac{V''}{V}\right)\left[
\frac{2\sigma}{2\sigma + V}\right]\label{eta}.
\end{eqnarray}
The modifications to General Relativity are contained in the square 
brackets of these expressions. They imply that {\it for a given 
potential and given initial conditions for the scalar field the 
slow--roll parameters are suppressed compared to the predictions made 
in General Relativity.} In other words, {\it brane world effects 
ease slow--roll inflation} \cite{wandsinflation}. 
In the limit $\sigma \ll V$ the parameter 
are heavily suppressed. It implies that steeper potentials can be used 
to drive slow--roll inflation \cite{copeland}. Let us discuss the implications for 
cosmological perturbations. 

According to Einstein's equation (\ref{einsteinbrane}), perturbations 
in the metric are sourced not only by matter perturbations {\it but also 
by perturbations of the bulk geometry}, encoded in the perturbation 
of $E_{ab}$. This term can be seen as an {\it external source} for 
perturbations, absent in General Relativity. If one regards $E_{ab}$
as an energy--momentum tensor of an additional fluid (called the Weyl-fluid), 
its evolution is connected to the energy density 
of matter on the brane, as one can see from eq. (\ref{weylfluid}). If one neglects the 
anisotropic stress of the Weyl-fluid, then at low energy and 
superhorizon scales, it decays as radiation, 
i.e. $\delta E_{ab} \propto a^{-4}$. However, 
the bulk gravitational field {\it exerts} an anisotropic stress onto the 
brane, whose  time-evolution cannot be obtained by considering 
the projected equations on the brane alone \cite{roy}. Rather, the full 
five--dimensional equations have to be solved, together with 
the junction conditions. The full evolution of $E_{ab}$ in the 
different cosmological eras is currently not 
understood. However, as we will discuss below, 
partial results have been obtained for the case 
of a de Sitter brane, which suggest that $E_{ab}$ does not change 
the spectrum of scalar perturbations. It should be noted however, that 
the issue is not settled and that it is also not clear if the subsequent 
cosmological evolution during radiation and matter era leaves an 
imprint of the bulk gravitational field in the anisotropies of the 
microwave background radiation \cite{rsperturbations}. 
With this in mind, we will, for scalar perturbations, neglect 
the gravitational backreaction described by the projected Weyl tensor. 

Considering scalar perturbations for the moment, the perturbed line 
element on the brane has the form
\begin{equation}
ds^2 = -(1 + 2A)dt^2 + 2\partial_{i}B dtdx^i 
+ ((1-2\psi)\delta_{ij} + D_{ij}E)dx^i dx^j,
\end{equation}
where the functions $A$, $B$, $E$ and $\psi$ depend on $t$ and 
$x^i$.

An elegant way of discussing  scalar perturbations is to make use of
of the gauge invariant quantity \cite{bardeen}
\begin{equation}\label{zeta}
\zeta = \psi + H \frac{\delta \rho}{\dot \rho}.
\end{equation}
In General Relativity, the evolution equation for $\zeta$ 
can be obtained from the energy--conservation equation \cite{malik}. 
It reads, on large scales,
\begin{equation}\label{zetaevolution}
\dot\zeta = -\frac{H}{\rho+p}\delta p_{\rm nad},
\end{equation}
where $\delta p_{\rm nad} = \delta p_{tot} - c_s^2\delta \rho$ 
is the non-adiabatic pressure perturbation.
The energy conservation equation, however, holds for the 
Randall--Sundrum model as well. Therefore, eq. (\ref{zetaevolution}) 
is still valid for the brane world model we consider. For inflation 
driven by a single scalar field $\delta p_{\rm nad}$ vanishes and therefore 
$\zeta$ is constant on superhorizon scales during inflation. 
Its amplitude is given in terms of the fluctuations in the scalar
field on spatially flat hypersurfaces:
\begin{equation}\label{zetafield}
\zeta = \frac{H\delta \phi}{\dot\phi}
\end{equation}
The quantum fluctuation in the (slow--rolling) scalar field obey 
$\langle (\delta\phi)^2 \rangle \approx (H/2\pi)^2$, as  the
Klein--Gordon equation is not modified in the brane world model we 
consider. The amplitude of scalar perturbations is \cite{liddlebook} 
$A_S^2 = 4 \langle \zeta^2 \rangle/25$. Using the slow--roll equations 
and  eq. (\ref{zetafield}) one obtains \cite{wandsinflation}
\begin{equation}\label{scalaramplitude}
A_S^2 \approx \left( \frac{512 \pi}{75 M_{pl}^6} \right) 
\frac{V^3}{V'^2} \left[\frac{2\sigma + V}{2\sigma}\right]^3 
|_{k=aH}
\end{equation}
Again, the corrections are contained in the terms in the square
brackets. For a given potential the amplitude of scalar perturbations 
is {\it enhanced} compared to the prediction of General Relativity. 

The arguments presented so far suggest that, at least for scalar
perturbations, perturbations in the bulk spacetime are not important 
during inflation. This, however, might not be true for tensor
perturbations, as gravitational waves can propagate into the bulk. 
For tensor perturbations, a wave equations for a single variable 
can be derived \cite{langloisperturbations}. 
The wave equation can be separated into a four--dimensional and a 
five--dimensonal part, so that the solution has the form 
$h_{ij} = A(y) h(x^\mu) e_{ij}$, where $e_{ij}$ is a (constant) 
polarization tensor. One finds that the amplitude for the zero mode of 
tensor perturbation is given by \cite{langloisperturbations}
\begin{equation}\label{tensoramplitude}
A_T^2 = \frac{4}{25 \pi M_{pl}^4} H^2 F^2(H/\mu)|_{k=aH},
\end{equation}
with
\begin{equation}
F(x) = \left[ \sqrt{1+x^2} - x^2 \sinh^{-1}\left(\frac{1}{x}\right)
\right]^{-1/2},
\end{equation}
where we have defined 
\begin{equation}
\frac{H}{\mu} = \left( \frac{3}{4\pi\sigma} \right)^{1/2} H M_{pl}.
\end{equation}

It can be shown that modes with $m>3H/2$ are generated but they 
decay during inflation. Thus, one expects in this scenario only the massless 
modes to survive until the end of inflation \cite{langloisperturbations}, 
\cite{rubakovperturbations}. 

From eqns. (\ref{tensoramplitude}) and (\ref{scalaramplitude}) one 
sees that the amplitudes of scalar and tensor perturbations are enhanced at 
high energies. However, 
scalar perturbations are more enhanced than tensors. Thus, the 
relative contribution of tensor perturbations will be suppressed, 
if inflation is driven at high energies.

Finally, we would like to mention that there are also differences between 
General Relativity and the brane world model we consider for the 
prediction of two--field brane inflation. Usually correlations are 
separated in  adiabatic and isocurvature modes for two--field inflation \cite{gordon}. 
In the Randall--Sundrum model, this correlation is suppressed if inflation 
is driven at high energies \cite{ashcroft}. This implies that isocurvature 
and adiabatic perturbations are uncorrelated, if inflation is driven at 
energies much larger than the brane tension. 

Coming back to cosmological perturbations, the biggest problem is that 
the evaluation of the projected Weyl tensor is only 
possible for the background cosmology. As soon as one tries to
analyse the brane cosmological perturbations, one faces the possibility
that the $E_{0i}$ terms might not vanish. 
In particular this means that the equation for the density contrast 
$\delta = \delta\rho/\rho$, 
which is given by ($w_m = p/\rho$, $k$ is the wavenumber)
\begin{equation}
\ddot \delta+ (2-3\omega_m)H\dot \delta -6\omega_m(H^2+\dot H)
\delta=(1+\omega_m)\delta R_{00}-\omega_m \frac{k^2}{a^2}\delta,
\end{equation}
cannot be solved as $\delta R_{00}$ involves $\delta E_{00}$
and can therefore not be deduced solely from the brane dynamics \cite{roy}. 

\subsection{Final Remarks on the Randall--Sundrum Scenario}
The Randall--Sundrum model discussed in this section is the simplest brane world 
model. We have not discussed other important conclusions one can draw from the 
modifications of Friedmann's equation, such as the evolution of primordial 
black holes \cite{raf}, its connection to the AdS/CFT correspondence 
(see e.g. \cite{gubser}-\cite{kaloper2}) and inflation driven by the trace 
anomaly of the conformal field theory living on the brane 
(see e.g. \cite{hawking1}-\cite{tracelast}). These developments are 
important in many respects, because they give not only insights about 
the early universe but  gravity itself. They will not be 
reviewed here. 

\section{Including a Bulk Scalar Field}
In this section we are going to generalize the previous results
obtained with an empty bulk. To be specific, we will consider the inclusion of a
scalar field in the bulk. As we will see, one can extend the projective 
approach wherein  one focuses on the
dynamics of the brane, i.e. one studies the projected Einstein {\it
and} the 
Klein-Gordon equation \cite{maedawands}, \cite{mennim}. As in the Randall-Sundrum setting, the
dynamics do not closed, as bulk effects do not decouple.
We will see that there are now two objects representing the bulk back-reaction:
the projected Weyl tensor $E_{\mu\nu}$ and the loss parameter
$\Delta\Phi_2$. In the case of homogeneous and isotropic
cosmology on the brane, the projected Weyl tensor is determined
entirely up to a dark radiation term. Unfortunately, no
information on the loss parameter is available. This prevents a
rigorous treatment of brane cosmology in the projective approach.

Another route amounts to studying the motion of a brane in a bulk
space-time. This approach is successful in the Randall-Sundrum
case thanks to Birkhoff's theorem which dictates a unique form for the metric in
the bulk \cite{charmousis}. In the case of a bulk scalar field, no such theorem is available. 
One has to resort to various ansatze for particular classes of bulk and 
brane scalar potentials (see e.g. \cite{kanti1}--\cite{kanti2}). 
We will come back to this in section 4.

\subsection{BPS Backgrounds}
\subsubsection{Properties of BPS Backgrounds}
As the physics of branes with bulk scalar fields is pretty
complicated, we will start with a particular example where both
the bulk and the brane dynamics are fully under control \cite{brax1} (see 
also \cite{groje} and \cite{youm}).
We specify the bulk Lagrangian as
\begin{equation}
S=\frac{1}{2\kappa_5^2}\int d^5x \sqrt{-g_5}\left(R-\frac{3}{4}\left(\left(\partial \phi\right)^2 
+V(\phi)\right)\right)
\end{equation}
where $V(\phi)$ is the bulk potential.
The boundary action depends on a brane potential $U_B(\phi)$
\begin{equation}
S_B=-\frac{3}{2\kappa_5^2}\int d^4x \sqrt{-g_4}U_B(\phi_0)
\end{equation}
where
$U_B(\phi_0)$ is evaluated on the brane.
The BPS backgrounds arise as particular case of this general
setting with a particular relationship between the bulk and brane potentials.
This relation appears in the study of $N=2$ supergravity  with
vector multiplets in the bulk. 
The bulk potential is given by
\begin{equation}
V=\left(\frac{\partial W}{\partial\phi}\right)^2-W^2
\end{equation}
where $W(\phi)$ is the superpotential. The brane potential
is simply given by the superpotential
\begin{equation}
U_B=W
\end{equation}
We would like to mention, that the last two relations have been also used 
in order to generate bulk solutions without necessarily imposing supersymmetry 
\cite{flanagan2},\cite{karch}.
Notice that the Randall-Sundrum case can be retrieved by putting $W=cst$.
Supergravity puts further constraints on the superpotential which
turns out to be of the exponential type \cite{brax1}
\begin{equation}
W=4k e^{\alpha \phi}
\end{equation}
with $\alpha=-1/\sqrt{12},1/\sqrt 3$.
In the following we will often choose this exponential potential
with an arbitrary $\alpha$ as an example.
The actual value of $\alpha$ does not play any role and will be
considered generic.

The bulk equations of motion comprise the Einstein equations and
the Klein-Gordon equation. In the BPS case and using the following ansatz 
for the metric
\begin{equation}\label{scalarfieldmetric}
ds^2 = a(y)^2 \eta_{\mu\nu}dx^\mu dx^\nu + dy^2,
\end{equation}
these second order differential equations reduce to a system of two 
first order differential equations 
\begin{eqnarray}
\frac{a'}{a}&=& -\frac{W}{4},\\\nonumber 
\phi'&=&\frac{\partial W}{\partial\phi}.\nonumber   
\end{eqnarray}
Notice that when $W=cst$ one easily retrieves the exponential
profile of the Randall-Sundrum model.

An interesting property of BPS systems can be deduced from the
study of the boundary conditions. The Israel junction conditions
reduce to
\begin{equation}
\frac{a'}{a}\vert_B= -\frac{W}{4}\vert_B
\end{equation}
and for the scalar field
\begin{equation}
\phi'\vert_B=\frac{\partial W}{\partial\phi}\vert_B  
\end{equation}
This is the main property of BPS systems: the boundary conditions
coincide with the bulk equations, i.e. as soon as the bulk
equations are solved one can locate the BPS branes anywhere in
this background, there is no obstruction due to the boundary conditions.
In particular two-brane systems with two boundary
BPS branes admit {\it moduli} corresponding to massless
deformations of the background. They are identified with the
positions of the branes in the BPS background. We will come back to 
this later in section 5.

Let us treat the example of the exponential superpotential.
The solution for the scale factor reads
\begin{equation}\label{aBPS}
a=(1-4k\alpha^2x_5)^{1/4\alpha^2},
\end{equation}
and the scalar field is given by 
\begin{equation}\label{phiBPS}
\phi=-\frac{1}{\alpha}\ln(1-4k\alpha^2x_5).
\end{equation}
For $\alpha\to 0$, the bulk scalar field decouples and these expressions 
reduce to the Randall-Sundrum case.
Notice a new feature here, namely the existence of singularities in the bulk,
corresponding to
\begin{equation}
a(x_5)\vert_{x_*}=0
\end{equation}
In order to analyse singularities it is convenient to use
conformal coordinates
\begin{equation}
du=\frac{dx_5}{a(x_5)}.
\end{equation}
In these coordinates light follows straight lines $u=\pm t$.
It is easy to see that
the singularities fall in two categories depending on $\alpha$.
For $\alpha^2< 1/4$ the singularity is at infinity $u_*=\infty$.
This singularity is null and absorbs incoming gravitons.
For $\alpha^2>1/4$ the singularity is at finite distance.
It is time-like and not wave-regular, i.e. the propagation of
wave packets is not uniquely defined in the vicinity of the singularity.
For all these reasons these naked singularities in the bulk are a
major drawback of brane models with bulk scalar fields \cite{braxsingularity}.
In the two-brane case the second brane has to sit in front of the
naked singularity.

\subsubsection{de Sitter and anti de Sitter Branes}
Let us modify slightly the BPS setting by detuning the
tension of the BPS brane. This corresponds to adding or
substracting some tension compared to the BPS case
\begin{equation}
U_B=TW
\end{equation}
where $T$ is real number.
Notice that this modification only affects the boundary
conditions, the bulk geometry and scalar field are still solutions
of the BPS equations of motion.
In this sort of situation, one can show that the brane does not stay
static.
For the detuned case, the result is either a boosted brane or a
rotated brane. We will soon generalize these results so we postpone
the detailed explanation to later.
Defining by $u(x^\mu)$ the position of the brane in conformal
coordinates, one obtains
\begin{equation}
(\partial u)^2= \frac{1-T^2}{T^2}.
\end{equation}
The brane velocity vector $\partial_\mu u$ is of constant norm.
For $T>1$, the brane velocity vector is time-like and the brane
moves at constant speed. For $T<1$ the brane velocity vector is
space-like and the brane is rotated.
Going back to a static brane,  we see that the bulk geometry and
scalar field become $x^\mu$ dependent.
In the next section we will find many more cases where branes
move in a static bulk or equivalently, a static brane borders a
non-static bulk.

Let us now conclude this section by studying the brane geometry
when $T>1$. In particular one can study the Friedmann equation for
the induced bulk factor
\begin{equation}
H^2=\frac{T^2-1}{16}W^2,
\end{equation}
where $W$ is evaluated on the brane. Of course we obtain  the
fact that cosmological solutions are only valid for $T>1$.
Now in the Randall-Sundrum case $W=4k$
leading to
\begin{equation}
H^2=(T^2-1)k^2.
\end{equation}
In the case $T>1$ the brane geometry is driven by a positive
cosmological constant. This is a de Sitter brane. When $T<1$ the
cosmological constant is negative, corresponding to an AdS brane.
We are going to generalize these results by considering the
projective approach to the brane dynamics.

\subsection{Bulk Scalar Fields and the Projective Approach}
\subsubsection{The Friedmann Equation}
We will first follow the traditional coordinate dependent path.
This will allow us to derive the matter conservation equation,
the Klein-Gordon and the Friedmann equations on the brane.
Then we will concentrate on the more geometric formulation where the
role of the projected Weyl tensor will become transparent 
\cite{brax2},\cite{vandebruck2}. Again, in this subsection we will put 
$\kappa_5\equiv1$.

We consider a static brane that we choose to put at the origin $x_5=0$.
and impose $ b(0,t)=1$.
This guarantees that the brane and bulk expansion rates
\begin{equation}
4H=\partial_\tau \sqrt{-g}\vert_0,\ 3H_B=\partial_\tau \sqrt{-g_4}\vert_0
\end{equation}
coincide. We have identified the brane cosmic time
$d\tau =ab\vert_0 dt$.
We will denote by prime the normal derivative
$
\partial_n=\frac{1}{ab}\vert_0 \partial_{x_5}
$.
Moreover we now allow for some matter to be present on the brane
\begin{equation}
\tau ^\mu_\nu\  ^{matter}=(-\rho_m,p_m,p_m,p_m).
\end{equation}
The bulk energy-momentum tensor reads
\begin{equation}
T_{ab}= \frac{3}{4}\left(\partial_a\phi\partial_b\phi\right)
-\frac{3}{8}g_{ab}\left(\left(\partial\phi\right)^2+V\right).
\end{equation}
The total matter density and pressure on the brane are given by 
\begin{equation}
\rho= \rho_m+\frac{3}{2}U_B,\ p=p_m -\frac{3}{2}U_B.
\end{equation}
The boundary condition for the scalar field is unchanged by the
presence of matter on the brane.

The $(05)$ Einstein equation leads to matter conservation
\begin{equation}
\dot \rho_m=-3H(\rho_m+p_m).
\end{equation}
By restricting the $(55)$ component of the Einstein equations we obtain
\begin{equation}
H^2=\frac{\rho^2}{36}-\frac{2}{3}Q-\frac{1}{9}E +\frac{\cal \mu}{a^4}
\end{equation}
in units of $\kappa_5^2$.
The last term is the dark radiation, whose origin is similar to 
the Randall-Sundrum case.
The quantity $Q$ and $E$ satisfy the differential equations \cite{vandebruck1}
\begin{eqnarray}
\dot Q+4HQ&=&HT^5_5,\nonumber \\
\dot E + 4HE &=& -\rho T^0_5. \nonumber 
\end{eqnarray}
These equations can be easily integrated to give
\begin{equation}
H^2= \frac{\rho_m^2}{36}+\frac{U_B\rho_m}{12}-\frac{1}{16a^4}\int d\tau
\frac{da^4}{d\tau}(\dot\phi^2-2U)-\frac{1}{12a^4}\int d\tau a^4
\rho_m \frac{dU_B}{d\tau},
\end{equation}
up to a dark radiation term and we have identified
\begin{equation}
U=\frac{1}{2}\left(U_B^2-\left(\frac{\partial U_B}{\partial \phi}\right)^2+V\right).
\end{equation}
This is  the Friedmann equation for a brane coupled to a bulk
scalar field.
Notice that retarded effects springing from the whole history
of the brane and scalar field dynamics are present. In the
following section we will see that these retarded effects come 
from the projected Weyl tensor. They result from the exchange
between the brane and the bulk.
Notice, that Newton's constant depends on the value of the 
bulk scalar field evaluated on the brane ($\phi_0=\phi(t,y=0)$):
\begin{equation}
\frac{8\pi G_N(\phi_0)}{3}
=\frac{\kappa_5^2 U_B(\phi_0)}{12}.
\end{equation}
On cosmological scale, time variation of the scalar field induce
a time variations of Newton's constant. This is highly 
constrained experimentally \cite{chiba},\cite{uzan}, 
leading to tight restrictions on the time dependence 
of the scalar field.

To get a feeling of the physics involved in the Friedmann equation,
it is convenient to assume that the scalar field is evolving
slowly on the scale of the variation of the scale factor.
Neglecting the evolution of Newton's constant, the Friedmann
equation reduces to
\begin{equation}
H^2= \frac{8\pi G_N(\phi)}{3}\rho_m +\frac{U}{8}-\frac{\dot \phi^2}{16}
\end{equation}
Several comments are in order. First of all we have neglected the
contribution due to the $\rho_m^2$ term as we are considering
energy scales below the brane tension. Then the main effect of
the scalar field dynamics is to involve the  potential energy
$U$ and the kinetic energy $\dot \phi^2$. Although the potential
energy appears with a positive sign we find that the kinetic
energy has a negative sign. 
For an observer on the brane this looks like a violation of unitarity.
We will reanalyse this issue in section 5, when considering the low energy
effective action in four dimensions and we will see that there is no unitarity problem
in this theory. The minus sign for the kinetic energy is due to
the fact that one does not work in the Einstein frame where
Newton's constant does not vary,
a similar minus sign appears also in the effective four--dimensional theory when
working in the brane frame.

The time dependence of the scalar field is determined by the
Klein-Gordon equation.
The dynamics is completely specified by 
\begin{equation}
\ddot\phi
+4H\dot\phi+\frac{1}{2}(\frac{1}{3}-\omega_m)\rho_m\frac{\partial U_B}{\partial\phi}=
-\frac{\partial U}{\partial\phi} +\Delta\Phi_2,
\end{equation}
where $p_m=\omega_m\rho_m$.
We have identified
\begin{equation}
\Delta\Phi_2=\phi''\vert_0-\frac{\partial U_B}{\partial
\phi}\frac{\partial^2 U_B}{\partial \phi^2}\vert_0.
\end{equation}
This cannot be set to zero and requires the knowledge of the
scalar field in the vicinity of the brane. When we discuss 
cosmological solutions below, we will assume that this term 
is negligible.

The evolution of the scalar field is driven by two effects. First
of all, the scalar field couples to the trace of the energy 
momentum tensor via the gradient of $U_B$. Secondly, the field is 
driven by the gradient of the potential $U$, which might not 
necessarily vanish. 

\subsubsection{The Friedmann equation vs the projected Weyl tensor}

We are now coming back to the origin of the non-trivial Friedmann equation.
Using the Gauss-Codazzi equation one can obtain the Einstein
equation on the brane \cite{maedawands},\cite{mennim}
\begin{equation}
\bar G_{ab}= -\frac{3}{8}Uh_{ab}+\frac{U_B}{4}\tau_{ab}+\pi_{ab}+\frac{1}{2}
\partial_a \phi \partial_b\phi -\frac{5}{16}(\partial\phi)^2 h_{ab}-E_{ab}.
\end{equation}
Now the projected Weyl tensor can be determined in the
homogeneous and isotropic cosmology case. Indeed  only the $E_{00}$
component is independent. Using the Bianchi identity $\bar D^a
\bar G_{ab}=0$ where $\bar D_a$ is the brane covariant
derivative, one obtains that
\begin{equation}
\dot E_{00}+4HE_{00}= \partial_{\tau}\left(\frac{3}{16}\dot \phi^2+\frac{3}{8}U\right)+
\frac{3}{2}H\dot\phi^2 +\frac{\dot U_B}{4}\rho_m
\end{equation}
leading to
\begin{equation}
E_{00}=\frac{1}{a^4}\int d\tau a^4\left(\partial_{\tau}\left(\frac{3}{16}\dot \phi^2
+\frac{3}{8}U\right) + \frac{3}{2}H\dot\phi^2 +\frac{\dot U_B}{4}\rho_m\right)
\end{equation}
Upon using
\begin{equation}
\bar G_{00}=3H^2
\end{equation}
one obtains the Friedmann equation.
It is remarkable that the retarded effects in the Friedmann
equation all spring from the projected Weyl tensor.
Hence the projected Weyl tensor proves to be much richer in the case
of a bulk scalar field than in the empty bulk case.

\subsubsection{Self-Tuning and Accelerated Cosmology}
The dynamics of the brane is not 
closed, it is an open system continuously exchanging energy with
the bulk. This exchange is characterized by the dark radiation
term and the loss parameter. Both require a detailed knowledge
of the bulk dynamics. This is of course beyond the projective
approach where only quantities on the brane are evaluated.
In the following we will {\it assume} that the dark radiation term
is absent and that the loss parameter is negligible. Furthermore, 
we will be interested in the effects of a bulk scalar field for 
late--time cosmology (i.e. well after nucleosynthesis) and not 
in the case for inflation driven by a bulk scalar field (see e.g. 
\cite{bscosfirst}-\cite{bscoslast}).

Let us consider the self-tuned scenario as a solution to the
cosmological constant problem. It corresponds to the BPS
superpotential with $\alpha=1$. In that case the potential $U=0$
for any value of the brane tension. 
The potential $U=0$ can be interpreted as a vanishing of the
brane cosmological constant. 
The physical interpretation of the vanishing of the cosmological constant
is that the brane tension curves the fifth dimensional space-time
leaving a flat brane intact.
Unfortunately, the description of the bulk geometry in that case
has shown that there was a bulk singularity which needs to be
hidden by a second brane whose tension is fine-tuned with
the first brane tension. This reintroduces a fine-tuning in the putative
solution to the cosmological constant problem \cite{nilles}.

Let us generalize the selftuned case to $\alpha\ne 1$, i.e. 
$U_B=TW,\  T>1$
and $W$ is the exponential superpotential. 
The resulting
induced metric on the brane is of the FRW type with a scale factor
\begin{equation}
a(t)= a_0 \left(\frac{t}{t_0}\right)^{1/3+ 1/6\alpha^2}
\end{equation}
leading to an acceleration parameter
\begin{equation}
q_0= \frac{6\alpha^2}{1+2\alpha^2}-1
\end{equation}
For the supergravity value $\alpha=-\frac{1}{\sqrt 12}$
this leads to $q_0=-4/7$. This is in  coincidental agreement with
the supernovae results. This model can serve as a {\it brane
quintessence model} \cite{brax1},\cite{brax2}. 
We will comment on the drawbacks of this model later. See also 
\cite{kiritsis} and \cite{tetradis} for similar ideas. 

\subsubsection{The brane cosmological eras}
Let us now consider the possible cosmological scenarios
with a bulk scalar field \cite{brax2},\cite{vandebruck2}. 
We assume  that the potential energy of the scalar field $U$ 
is negligible throughout the radiation and matter eras before 
serving as quintessence in the recent past.

At very high energy above the tension of the brane 
the non-conventional cosmology driven by the $\rho_m^2$ term in
the Friedmann equation is obtained. Assuming radiation domination, 
the scale factor behaves like
\begin{equation}
a=a_0\left(\frac{t}{t_0}\right)^{1/4}
\end{equation}
and the scalar field
\begin{equation}
\phi=\phi_i+ \beta \ln\left(\frac{t}{t_0}\right)
\end{equation}
In the radiation dominated era, no modification is present, provided
\begin{equation}
\phi=\phi_i
\end{equation}
which is a solution of the Klein-Gordon equation as the trace of
the energy-momentum of radiation vanishes (together with a decaying 
solution, which we have neglected).
In the matter dominated era the scalar field
evolves due to the coupling to the trace of the energy-momentum tensor.
This has two consequences. Firstly, the kinetic energy of the
scalar field starts contributing in the Friedmann equation.
Secondly, the effective Newton constant does not remain constant.
The cosmological evolution of Newton's constant is severely
constrained since nucleosynthesis \cite{chiba},\cite{uzan}. 
This restricts the possible time variation of $\phi$.

In order to be more quantitative let us come back to the
exponential superpotential case with a detuning parameter $T$. 
The time dependence of the scalar field and
scale factor become
\begin{eqnarray}
\phi&=&\phi_1-\frac{8}{15}\alpha \ln\left(\frac{t}{t_e}\right)\nonumber \\  
a&=& a_e\left(\frac{t}{t_e}\right)^{\frac{2}{3}-\frac{8}{45}\alpha^2}\nonumber 
\end{eqnarray}
where $t_e$ and $a_e$ are the time and scale factor at
matter-radiation equality.
Notice the slight discrepancy of the scale factor exponent with
the standard model value of $2/3$.
The redshift dependence of the Newton constant is
\begin{equation}
\frac{G_N(z)}{G_N(z_e)}=\left(\frac{z+1}{z_e+1}\right)^{4\alpha^2/5}
\end{equation}
For the supergravity model with $\alpha=-\frac{1}{\sqrt 12}$ and
$z_e\sim 10^3$ this leads to a decrease by (roughly) 37\% since 
nucleosynthesis. This is marginally compatible with experiments 
\cite{chiba},\cite{uzan}.

Finally let us analyse the possibility of using the brane
potential energy of the scalar field $U$ as the source of
acceleration now. We have seen that when matter is negligible on
the brane, one can build brane quintessence models. We now require
that this occurs only 
in the recent past. As can be expected, this leads to a fine-tuning
problem as 
\begin{equation}
M^4\sim \rho_c
\end{equation}
where $M^4=(T-1)\frac{3W}{2\kappa_5^2}$
is the amount of detuned tension on the brane.
Of course this is nothing but a reformulation of the usual
cosmological constant problem. Provided one accepts this
fine-tuning, as in most quintessence models, 
the exponential  model with $\alpha=-\frac{1}{\sqrt 12}$is a
cosmological consistent quintessence model with a five dimensional origin.

\subsection{Brief summary}
The main difference between a brane world model with a bulk scalar field 
and the Randall--Sundrum model is that the gravitational constant becomes 
time--dependent. As such it has much in common with scalar--tensor theories 
\cite{fujiibook}, but there are important differences due to the 
projected Weyl tensor $E_{\mu\nu}$ and its time--evolution.   
The bulk scalar field can play the role of the quintessence 
field, as discussed above, but  it could also play a role in an inflationary 
era in the very early universe (see e.g. \cite{bscosfirst}-\cite{bscoslast}). 
In any case, the cosmology of such a system is much richer and, because of the 
variation of the gravitational constant, more constrained. It
remains to be seen
if the bulk scalar field can leave a trace in the CMB anisotropies and 
Large Scale Structures (for first results see \cite{vandebruck2}).

\section{Moving Branes in a Static Bulk}
So far, we were mostly concerned with the evolution of the brane, 
without referring to the bulk itself. In fact, the coordinates introduced 
in eq. (\ref{metricbraneframe}) are a convenient choice for  studying the brane 
itself, but when it comes to analysing the bulk dynamics and its geometry, 
these coordinates are not the best choice. We have already mentioned the 
extended Birkhoff theorem in section 2. It states that for the case of a vacuum 
bulk spacetime, the bulk is necessarily static, in certain coordinates. A 
cosmological evolving brane is then moving in that spacetime, whereas for an 
observer confined on the brane the motion of the brane will be seen as an 
expanding (or contracting) universe. In the case of a scalar field in the bulk, 
a similar theorem is unfortunately not available, which makes the study of such 
systems much more complicated. We will now discuss these issues in some detail, 
following in particular \cite{stephen1} and \cite{stephen2}.

\subsection{Motion in AdS-Schwarzschild Bulk} 

We have already discussed the static background associated with 
BPS configurations (including the Randall--Sundrum case) in the last 
section. Here we will focus  on other backgrounds for 
which one can integrate the bulk equations of motion. 
Let us write the following ansatz for the metric 
\begin{equation}\label{staticcoordinates}
ds^2 = -A^2(r)dt^2 + B^2(r) dr^2 + R^2(r)d\Sigma^2
\end{equation}
where $d\Sigma^2$ is the metric on the 3d symmetric space of
curvature $q=0,\pm 1$. In general, the function $A,\ B$ and $R$ 
depend on the type of scalar field potential. 
This is to be contrasted with the case of a negative bulk
cosmological constant where Birkhoff's theorem states that the 
most general solution of the (bulk) Einstein equations is given  
by $A^2=f$, $B^2=1/f$ and $R=r$ where 
\begin{equation}
f(r)=q+\frac{r^2}{l^2}-\frac{\mu}{r^2}. 
\end{equation}
We have denoted by $l=1/k=\sqrt{-6/(\Lambda_5\kappa_5^2)}$  
the AdS scale and $\mu$ the black hole mass (see section 2). 
This solution is the so--called AdS-Schwarzschild solution.

Let us now study the motion of a brane of tension $T/l$ in such a background. 
The equation of motion is determined by the junction conditions. 
The method will be reviewed later when a scalar field is present 
in the bulk. The resulting equation of motion for a boundary 
brane with a $Z_2$ symmetry is 
\begin{equation} 
\left(\dot r^2+f(r)\right)^{1/2}=\frac{T}{l}r
\end{equation}
for a brane located at $r$ \cite{kraus}. Here $\dot r$ is the velocity of the
brane measured with the proper time on the brane.
This leads to the following Friedmann equation
\begin{equation}
H^2 \equiv \left(\frac{\dot r}{r}\right)^2 
= \frac{T^2-1}{l^2}-\frac{q}{r^2}+\frac{\mu}{r^4}.
\end{equation}
So the brane tension leads to an effective cosmological constant
$(T^2-1)/l^2$. The curvature gives the usual term familiar from
standard cosmology while the last term is the dark radiation term whose
origin springs from the presence of a black-hole in the bulk.
At late time the dark radiation term is negligible for an
expanding universe, we retrieve the cosmology of a FRW universe
with a non-vanishing cosmological constant. The case $T=1$
corresponds of course to the Randall-Sundrum case.

\subsection{Moving branes}
Let us now describe the general formalism, which covers 
the case of the AdS--Schwarzschild spacetime mentioned above.
 
Consider a brane embedded in a static background. It is 
parametrized by the coordinates $X^A(x^\mu)$ where $A=0\dots 4$ 
and the $x^\mu$ are world volume coordinates.
Locally the brane is characterized by the local frame
\begin{equation}
e^A_{\mu}=\frac{\partial X^A}{\partial x^\mu},
\end{equation}
which are tangent to the brane. The induced metric is given by
\begin{equation}
h_{\mu\nu}= g_{AB}e^A_\mu e^B_\nu
\end{equation}
and the extrinsic curvature
\begin{equation}
K_{\mu\nu}
=e^A_\mu e^B_\nu D_A n_B,
\end{equation}
where $n^A$ is the unit vector normal to the brane defined by (up
to a sign ambiguity)
\begin{equation}
g_{AB}n^An^B=1,\ n_Ae^A_{\mu}=0.
\end{equation}
For a homogeneous brane embedded in the spacetime described by
the metric (\ref{staticcoordinates}), we have $T=T(\tau)$, $r=r(\tau)$ 
where $\tau$ is the proper time on the brane.
The induced metric is
\begin{equation}
ds_B^2=-d\tau^2 +R^2(\tau) d\Sigma^2.
\end{equation}
The local frame becomes
\begin{equation}
e^A_{\tau}=(\dot T,\dot r, 0,0,0),\ e^A_i=(0,0,\delta^A_i),
\end{equation}
while the normal vector reads
\begin{equation}
n_A=(AB \dot r, -B\sqrt{1+\dot r^2},0,0,0).
\end{equation}
The components of the extrinsic curvature tensor can found to be
\begin{eqnarray}
K_{ij}&=&-\frac{\sqrt{1+B^2\dot r^2}}{B}RR'\delta_{ij}, \\
K_{\tau\tau}&=&\frac{1}{AB}\frac{d}{dr}(A\sqrt{1+B^2\dot r^2}).
\end{eqnarray}
The junction conditions are given by 
\begin{equation}
K_{\mu\nu}=-\frac{\kappa_5^2}{2}\left(\tau_{\mu\nu}-\frac{1}{3}\tau h_{\mu\nu}\right).
\end{equation}
This implies that the brane dynamics are specified by the equations of motion
\begin{equation}
\frac{\sqrt{1+B^2\dot r^2}}{B}\frac{R'}{R}=\frac{\kappa_5^2}{6}\rho
\end{equation}
and
\begin{equation}
\frac{1}{AB}\frac{d}{dr}(A\sqrt{1+B^2\dot r^2})=-\frac{\kappa_5^2}{6}(2\rho+3p),
\end{equation}
where we have assumed a fluid description for the matter on the brane. 
These two equations determine the dynamics of any brane in a
static background.

Let us now close the system of equations  by stating the scalar field
boundary condition \cite{braxlanglois}
\begin{equation}
n^A\partial_A\phi=\frac{\kappa_5^2}{2}\frac{d\xi}{d\phi}(\rho-3p),
\end{equation}
where the coupling to the brane is defined by the Lagrangian
\begin{equation}
S_{brane}=\int d^4 x {\cal L}[\psi_m,\tilde h_{\mu\nu}],
\end{equation}
where $\psi_m$ represents the matter fields and
\begin{equation}
\tilde h_{\mu\nu}=e^{2\xi (\phi)}h_{\mu\nu}.
\end{equation}
This reduces to
\begin{equation}
\phi'=\frac{\kappa_5^2}{2}\frac{B}{\sqrt{1 + 
B^2 \dot r^2}}\frac{d\xi}{d\phi}(-\rho +3p).
\end{equation}
Combining the junction conditions leads to the  conservation equation 
\begin{equation} 
\dot \rho + 3H(\rho+ p)= (\rho-3p)\dot \xi.
\end{equation}
This is nothing but the conservation of matter in the Jordan 
frame defined by $\tilde h_{\mu\nu}$. 

We now turn to a general analysis of the brane motion in a
static bulk. To do that it is convenient to parametrized the bulk
metric slightly differently
\begin{equation}
ds^2= -f^2(r)h(r) dt^2 +\frac{dr^2}{h(r)}+ r^2d\Sigma^2.
\end{equation}
Now, the Einstein equations lead to (redefining $\phi \to
\frac{\sqrt 3}{2\kappa_5}\phi$ and $V\to \frac{3}{8\kappa_5^2} V$)
\begin{eqnarray}
& &\frac{3}{r^2}\left(h+\frac{rh'}{2}-q\right) 
= -{\kappa_5^2}\left( \frac{h\phi'^2}{2}+V\right) \label{eins}\\ 
& &\frac{3}{r^2}\left(h+\frac{rh'}{2}-q+\frac{hrf'}{f}\right) 
= {\kappa_5^2}\left(\frac{h\phi'^2}{2}-V\right) \label{zwei}
\end{eqnarray}
and the Klein Gordon equation
\begin{equation}
h\phi''+\left(\frac{3h}{r}+\frac{h f'}{f}+h'\right)\phi'=\frac{dV}{d\phi}.
\end{equation}
Subtracting eq. (\ref{eins}) from (\ref{zwei}) and solving the resulting 
differential equation, we obtain 
\begin{equation}
f=\exp\left(\frac{\kappa_5^2}{3}\int dr r\phi'^2\right).
\end{equation}
It is convenient to evaluate the spatial trace of the projected
Weyl tensor. This is obtained by computing both the bulk Weyl
tensor and the vector normal to the moving brane. With 
$A=\sqrt h f,\ R=r,\ B= 1/\sqrt h$, this gives 
\begin{equation}
\frac{\mu}{r^4}=-\frac{E^i_i}{3}=\frac{r}{4f^2}\left(\frac{h f^2}{r^2}\right)'+\frac{q}{2r^2}.
\end{equation}
This is the analogue of the dark radiation term for a general background. 
The equations of motion can be cast in the form 
\begin{eqnarray}
\mu'&=&-\frac{\kappa_5^2}{3}(\mu-\frac{kr^2}{2})r\phi'^2, \\
{\cal H}'+4\frac{\mu}{r^5}&=&-\frac{2\kappa_5^2}{3}({\cal H}-\frac{q}{r^2})r\phi'^2, \\
\kappa_5^2 V&=&6{\cal H}+\frac{3}{4}r{\cal H}'-3\frac{\mu}{r^4},
\end{eqnarray}
where we have defined
\begin{equation}
{\cal H}=\frac{q-h}{r^2}.
\end{equation}
This allows to retrieve easily some of the previous solutions. 
Choosing $\phi$ to be constant leads to $f=1$, $\mu$ is constant and 
\begin{equation}
{\cal H}=-\frac{1}{l^2}+\frac{\mu}{r^4}
\end{equation}
This is the AdS-Schwarzschild solution. 

For $q=0$ the equations of motion simplify to
\begin{eqnarray} 
\frac{\kappa_5^2}{3}r\phi'&=&-\frac{d\ln\mu}{d\phi}, \label{togetr} \\
d\left(\frac{\cal H}{\mu^2}\right)&=&\frac{1}{\mu}d\left(r^{-4}\right), \\
\frac{\kappa_5^2}{6}V&=&-\frac{3}{4\kappa_5^2}\frac{d\mu}{d\phi}\frac{d}{d\phi}
\left(\frac{\cal H}{\mu}\right) + {\cal H}\label{V}
\end{eqnarray}
In this form it is easy to see that the dynamics of the bulk are
completely integrable.
First of all the solutions depend on an arbitrary function $\mu (\phi)$
which determines the dynamics.
Notice that 
\begin{equation}
f=\frac{\mu_0}{\mu}
\end{equation}
where $\mu_0$ is an arbitrary constant. 
The radial coordinate $r$ is obtained by simple integration of eq. (\ref{togetr}) 
\begin{equation}
r=r_0e^{-\frac{\kappa_5^2}{3}\int \frac{d\phi}{d\ln\mu}d\phi}.
\end{equation}
Finally the rest of the metric follows from 
\begin{equation} 
h=-\frac{4\kappa_5^2}{3}r^2 \mu^2 \int d\phi \frac{d\phi}{d\mu}
e^{4\frac{\kappa_5^2}{3}\int \frac{d\phi}{d\ln\mu}d\phi}
\end{equation}
The potential $V$ then follows (\ref{V}).
This is remarkable and  shows why Birkhoff's theorem 
is not valid in the presence of a bulk scalar field.
Moreover, it is intriguing that the generalization of the dark
energy term dictates the bulk dynamics completely.

It is interesting to recast the Friedmann equation in the form
\begin{equation}
H^2= {\cal H}+ \frac{\kappa_5^4}{36}\mu^2\rho^2
\end{equation}
where $H$ is the Hubble parameter on the brane in cosmic time.
One can retrieve standard cosmology by studying the dynamics in
the vicinity of a critical point $\frac{d\mu}{d\phi}=0$.
Parametrizing
\begin{equation}
\mu= \frac{6A}{\kappa_5^2}+B\phi^2
\end{equation}
leads to the Friedmann equation
\begin{equation}
H^2=\frac{\kappa_5^4}{36}(\rho^2-\theta)\mu^2 +\frac{\mu}{a^4}+ o(a^{-4})
\end{equation}
Here $\theta$ is an arbitrary integration constant.
Notice that this is a small deviation from the Randall-Sundrum
case as
\begin{equation}
\phi= r^{-B/A}
\end{equation}
goes to zero at large distances. Hence, standard cosmology is
retrieved at low energy and long distance.

\section{Cosmology of a Two--Brane System}
In this section we will once more include an ingredient suggested 
by particle physics theories, in particular M--theory. So far we have 
assumed that there is only one brane in the whole space--time.
According to string theory, there should be at least  another brane in the bulk. 
Indeed, in heterotic M--theory these branes are the boundaries of the 
bulk spacetime \cite{horavawitten}. However, even from a purely phenomenological 
point of 
view there is a reason to include a second brane: the bulk singularity
(or the AdS horizon). As we have seen in section 3, the inclusion of a 
bulk scalar field often implies the presence of a naked
singularity located away from the positive tension brane. The second 
brane which we include now should shield this singularity, so that 
the physical spacetime stretches between the two branes. Another 
motivation is the hierarchy problem. Randall and Sundrum proposed 
a two brane model (one with positive and one with negative tension), 
embedded in a five--dimensional AdS spacetime. In their scenario the 
standard model particles would be confined on the {\it negative} 
tension brane. As they have shown, in this case gravity is weak
due to the warping of the bulk spacetime. However, as will become clear 
from the results in this section, in order for this model to be consistent 
with gravitational experiments, the interbrane distance has to be {\it fixed} 
\cite{garriga}. This can be achieved, for example, with  a bulk scalar field. 
As  shown in \cite{garriga} and \cite{chiba2}, gravity in the two--brane 
model of Randall-Sundrum is described by a scalar--tensor theory, in which the 
interbrane--distance, called radion, plays the role of a  scalar field. 
The bulk scalar field will modify the Brans--Dicke parameter (see \cite{brax3} and 
\cite{brax4}) of the scalar field and will introduce a second scalar field in the 
low--energy effective theory, so that the resulting theory at low energy in 
the case of two branes and a bulk scalar field is a bi--scalar--tensor theory 
\cite{vandebruck3},\cite{cynolter}. 

In the following we will investigate the cosmological consequences when
the distance between the branes is {\it not} fixed (for some aspects 
not covered here see e.g. \cite{radcosfirst}-\cite{radcoslast}). Motivation for this 
comes, for example, from a recent claim that the fine--structure constant 
might slowly evolve with time \cite{barrow}. 

\subsection{The low--energy effective action}
In order to understand the cosmology of the two--brane system, we
derive the low-energy effective action by utilizing the moduli space 
approximation. From the discussion in section 3 and section 4 it becomes 
clear, that the {\it general} solution of the bulk Einstein equations for
a given potential is difficult to find. The moduli space approximation 
gives the {\it low--energy--limit} effective action for the two brane 
system, i.e. for energies much smaller than the brane tensions. 

In the static BPS solutions described in the section 3, 
the brane positions can be chosen arbitrarily. In other words, they 
are {\it moduli fields}. It is expected that by putting some 
matter on the branes, these moduli field become time-dependent, or, 
if the matter is inhomogeneously distributed, space--time dependent. 
Thus, the first approximation is to replace the brane--positions 
with space--time dependent functions. Furthermore, in order to allow 
for the gravitational zero--mode, we will replace the flat spacetime 
metric $\eta_{\mu\nu}$ with $g_{\mu\nu}(x^\alpha)$. We do assume that
the evolution of these fields is slow, which means that we neglect 
terms like $(\partial \phi)^3$ when constructing the low-energy
effective action. 

As already mentioned, the moduli space approximation is only a good approximation 
at energies much less than the brane tension. {\it Thus, we do 
not recover the quadratic term in the moduli space approximation.} 
We are interested in the {\it late time} effects after
nucleosynthesis, where the corrections have to be small.

Replacing $\eta_{\mu\nu}$ 
with $g_{\mu\nu}(x^{\alpha})$ in (\ref{scalarfieldmetric}) and 
collecting all the terms one finds from the 5D action after an integration 
over $y$:
\begin{eqnarray}
S_{\rm MSA} &=& \int d^4 x \sqrt{-g_4}\left[ f(\phi,\sigma) {\cal R}^{(4)} 
+ \frac{3}{4}a^2(\phi)\frac{U_B(\phi)}{\kappa_5^2}(\partial \phi)^2 \right.
\nonumber \\
&-& \left. \frac{3}{4} a^2(\sigma)\frac{U_B}{\kappa_5^2}(\sigma)(\partial \sigma)^2 \right].
\end{eqnarray}
with 
\begin{equation}
f(\phi,\sigma) = \frac{1}{\kappa_5^2} \int^{\sigma}_{\phi} dy a^2 (y),
\end{equation}
with $a(y)$  given by (\ref{aBPS}).
The moduli $\phi$ and $\sigma$ represent the location
of the two branes.
Note that the kinetic term of the field $\phi$ has the wrong
sign. This is an artifact of the frame we use here. As we will see 
below, it is possible to go to the Einstein frame with a simple
conformal transformation, in which the sign in front of the kinetic
term is correct for both fields.

In the following we will concentrate on the BPS system with exponential 
superpotential from section 3. Let us redefine the fields according to 
\begin{equation} 
\tilde \phi^2 = \left(1 - 4k\alpha^2 \phi\right)^{2\beta}, \label{posia1}
\tilde \sigma^2 = \left(1-4k\alpha^2 \sigma\right)^{2\beta} \label{posia2},
\end{equation} 
with $ \beta = \frac{2\alpha^2 + 1}{4\alpha^2}$; 
and then
\begin{equation}
\tilde \phi = Q \cosh R, \label{posib1}  \
\tilde \sigma = Q \sinh R \label{posib2}.
\end{equation}
A conformal transformation $\tilde g_{\mu\nu} = Q^2 g_{\mu\nu}$ 
leads to the Einstein frame action:
\begin{eqnarray}
S_{\rm EF} &=& \frac{1}{2k\kappa^2_5(2\alpha^2 + 1)} 
\int d^4x \sqrt{-g}\left[ {\cal R} -  \frac{12\alpha^2}{1+2\alpha^2}
\frac{(\partial Q)^2}{Q^2} \right. \nonumber \\
&-& \left. \frac{6}{2\alpha^2 + 1}(\partial R)^2\right].
\end{eqnarray}
Note that in this frame both fields have the correct sign in front of
the kinetic terms. For $\alpha \rightarrow 0$ 
(i.e.\ the Randall--Sundrum case) the $Q$--field decouples. This 
reflects the fact, that the bulk scalar field decouples, and the only 
scalar degree of freedom is the distance between the branes. 
One can read off the gravitational constant to be 
\begin{equation}
16\pi G = 2k\kappa_5^2 (1+2\alpha^2).
\end{equation}

The matter sector of the action can be found easily: if matter lives
on the branes, it ``feels'' the induced metric. That is, the action 
has the form 
\begin{equation}
S_m^{(1)} = S_m^{(1)}(\Psi_1,g^{B(1)}_{\mu\nu}) \hspace{0.5cm} {\rm and}
\hspace{0.5cm} S_m^{(2)} = S_m^{(2)}(\Psi_2,g^{B(2)}_{\mu\nu}),
\end{equation}
where $g^{B(i)}_{\mu\nu}$ denotes the induced metric on each branes. 
In going to the Einstein frame one gets 
\begin{equation}
S_m^{(1)} = S_m^{(1)}(\Psi_1,A^2(Q,R)g_{\mu\nu}) \hspace{0.5cm} {\rm and}
\hspace{0.5cm} S_m^{(2)} = S_m^{(2)}(\Psi_2,B^2(Q,R)g_{\mu\nu}),
\end{equation}
where matter now couples explicitely to the fields via the functions
$A$ and $B$, which we will give below (neglecting derivative interactions).

The theory derived with the help of the moduli space approximation 
has the form of a {\it multi--scalar}--tensor theory, in which 
matter on both branes couple differently to the moduli fields. 
We note, that methods different from the moduli--space approximation 
have been used in the literature in order to obtain the low--energy 
effective action or the resulting field equations for a two--brane 
system (see in particular 
\cite{braneaction0}--\cite{braneaction4}). Qualitatively, the features 
of the resulting theories agree with the moduli--space approximation 
discussed above.

In the following we will discuss observational constraints imposed 
on the parameter of the theory. 

\subsection{Observational constraints}
In order to constrain the theory, it is convenient to write the moduli 
Lagrangian in the form
of a non-linear sigma model with kinetic terms
\begin{equation}
\gamma_{ij}\partial \phi^i\partial \phi^j,
\end{equation}
where $i=1,2$ labels the moduli $\phi^1=Q$ and $\phi^2=R$.
The sigma model couplings are here
\begin{equation}
\gamma_{QQ}= \frac{12\alpha^2}{1+2\alpha^2}\frac{1}{Q^2},\
\gamma_{RR}=\frac{6}{1+2\alpha^2}.
\end{equation}
Notice the potential danger of the $\alpha\to 0$ limit, the RS model,
where the coupling to $Q$ becomes very small. In an ordinary
Brans-Dicke theory with a single field, this would correspond to a
vanishing Brans-Dicke parameter which is ruled out
experimentally. Here we will see that the coupling to matter is such
that this is not the case. Indeed we can write the action expressing
the coupling to
ordinary matter on our brane 
as
\begin{equation}  
  A=a(\phi)f^{-1/2}(\phi,\sigma) , \  B=a(\sigma)f^{-1/2}(\phi,\sigma) ,
\end{equation}
where we have neglected the derivative interaction. 

Let us introduce the parameters
\begin{equation}
\alpha_Q=\partial_Q \ln A,\ \alpha_R=\partial_R \ln A.
\end{equation}
We find that ($\lambda = 4/(1+2\alpha^2)$) 
\begin{equation}
A=Q^{-\frac{\alpha^2\lambda}{2}}(\cosh R)^{\frac{\lambda}{4}},
\end{equation}
leading to
\begin{equation}
\alpha_Q= -\frac{\alpha^2\lambda}{2}\frac{1}{Q}, \alpha_R=\frac{\lambda
\tanh R}{4}.
\end{equation}
Observations constrain the parameter
\begin{equation}
\theta=\gamma^{ij}\alpha_i\alpha_j
\end{equation}
to be less than $10^{-3}$ \cite{gilles}. 
We obtain therefore a bound on
\begin{equation}
\theta= \frac{4}{3}\frac{\alpha^2}{1+2\alpha^2}+ \frac{\tanh^2 R}{6(1+2\alpha^2)}.
\end{equation}
The bound implies that
\begin{equation}
\alpha \leq 10^{-2},\ R\leq 0.2
\end{equation}
 The smallness of $\alpha$ indicates a strongly warped
bulk geometry such as an Anti--de Sitter spacetime.  In the case
$\alpha=0$, we can easily interpret the bound on $R$.  Indeed in that
case
\begin{equation}
\tanh R = e^{-k(\sigma -\phi)},
\end{equation}
i.e. this is nothing but the exponential of the radion field measuring
the distance between the branes. We obtain  that
gravity experiments require the branes to be sufficiently far apart.
When $\alpha\ne 0$ but small, one way of obtaining a small value of $R$
is for the hidden brane to become close from the would-be singularity
where $a(\sigma)=0$.

We would like to mention that the parameter $\theta$ can be calculated 
also for matter on the negative tension brane. Then, following the 
same calculations as above, it can be seen that the observational
constraint for $\theta$ {\it cannot} be satisfied. Thus, if the
standard model particles are confined on the negative tension brane, 
{\it the moduli have necessarily to be stabilized.} In the following 
we will assume that the standard model particles are confined on the 
positive tension brane and study the cosmological evolution of
the moduli fields. 

\subsection{Cosmological implications}
The discussion in the last subsection raises an important question: 
the parameter $\alpha$ has to be choosen rather small, in order for 
the theory to be consistent with observations. Similarly the field $R$ 
has to be small too. The field $R$ is dynamical and one would
like to know if the cosmological evolution drives the field $R$ to 
small values such that it is consistent with the observations today. 
Otherwise  are there natural initial conditions for the field 
$R$? In the following we study the cosmological evolution of the system in order to 
answer these questions. 

The field equations for a homogenous and isotropic universe can be
obtained from the action. The Friedmann 
equation reads
\begin{equation}\label{Friedmann}
H^2 = \frac{8 \pi G}{3} \left(\rho_1 + \rho_2 + V_{\rm eff} 
+ W_{\rm eff} \right) + \frac{2\alpha^2}{1 + 2\alpha^2} \dot\phi^2
+ \frac{1}{1+2\alpha^2} \dot R^2.
\end{equation}
where we have defined $Q = \exp \phi$. 
The field equations for $R$ and $\phi$ read
\begin{eqnarray}
\ddot R + 3 H \dot R &=& - 8 \pi G \frac{1+2\alpha^2}{6}\left[ 
\frac{\partial V_{\rm eff}}{\partial R} + 
\frac{\partial W_{\rm eff}}{\partial R} \right. \nonumber \\ 
&+& \left. \alpha_R^{(1)} (\rho_1 - 3p_1) + 
\alpha_R^{(2)} (\rho_2 - 3p_2) \right] \label{Rcos} 
\end{eqnarray} 
\begin{eqnarray}
\ddot \phi + 3 H \dot \phi &=&
-8 \pi G \frac{1+2\alpha^2}{12 \alpha^2} \left[ 
\frac{\partial V_{\rm eff}}{\partial \phi} + 
\frac{\partial W_{\rm eff}}{\partial \phi} \right. \nonumber \\
&+& \left. \alpha_\phi^{(1)} (\rho_1 - 3p_1) + 
\alpha_\phi^{(2)} (\rho_2 - 3p_2) \right].\label{Qcos}
\end{eqnarray}
The coupling parameter are given by
\begin{eqnarray}
\alpha_\phi^{(1)} &=& -\frac{2\alpha^2}{1+2\alpha^2}, \hspace{0.5cm}
\alpha_\phi^{(2)} = -\frac{2\alpha^2}{1+2\alpha^2}, \label{coupling1} \\
\alpha_R^{(1)} &=& \frac{\tanh R}{1+2\alpha^2}, \hspace{0.5cm}
\alpha_R^{(2)} = \frac{(\tanh R)^{-1}}{1+2\alpha^2}. \label{coupling2}
\end{eqnarray}
We have included matter on both branes as well as potentials 
$V_{\rm eff}$ and $W_{\rm eff}$ on each branes. 
We now concentrate on the case where matter is only on our brane. 
In the radiation dominated epoch the trace of the energy--momentum tensor vanishes, so that 
$Q$ and $\phi$ quickly become constant. The scale factor scales like 
$a(t) \propto t^{1/2}$.

In the matter--dominated era, the solution to these equations is given by 
\begin{equation}
\rho_1 = \rho_e\left(\frac{a}{a_e}\right)^{-3-2\alpha^2/3}
,a=a_e\left(\frac{t}{t_e}\right)^{2/3-4\alpha^2/27}
\end{equation}
together with
\begin{eqnarray}
\phi &=& \phi_e+\frac{1}{3}\ln\frac{a}{a_e}, {\mbox{\vspace{0.5cm}}} 
R = R_0\left(\frac{t}{t_e}\right)^{-1/3}
+ R_1\left(\frac{t}{t_e}\right)^{-2/3},
\end{eqnarray}
as soon as $t\gg t_e$. Note that $R$ indeed decays. This implies
that small values of $R$ compatible with gravitational
experiments are favoured by the cosmological evolution. Note, however, 
that the size of $R$ in the early universe is constrained by 
nucleosynthesis as well as by the CMB anisotropies.
A large discrepancy between the values of $R$ during
nucleosynthesis and now induces a variation of the particle
masses, or equivalently Newton's constant, which is excluded experimentally. 
One can show that 
by putting matter on the negative tension brane as well, the 
field $R$ evolves even faster to zero \cite{vandebruck3}. This behaviour is reminiscent 
of the attractor solution in scalar--tensor theories \cite{damour}.

In the five--dimensional picture the fact that $R$ is driven to small 
values means that the negative tension brane is driven towards the 
bulk singularity. In fact, solving the equations numerically for more general
cases suggest that $R$ can even by negative, which is, in the 
five--dimensional description meaningless, as the negative tension 
brane would move through the bulk singularity.
Thus, in order to make any further 
progress, one has to understand the bulk singularity
better\footnote{For $\alpha=0$ the theory is equivalent to 
the Randall--Sundrum model. In this case the bulk singularity is
shifted towards the Anti--de Sitter boundary.}. Of
course, one could simply assume that the negative tension brane 
is destroyed when it hits the singularity. A more interesting
alternative would be if the brane is repelled instead. It was
speculated that this could be described by some effective potential 
in the low-energy effective action \cite{vandebruck3}.

\section{Epilogue: Brane Collision}
We have seen that brane world models are plagued with a
singularity problem: the negative tension brane might hit a bulk 
singularity. In that case our description of the physics on the
brane requires techniques beyond the field theory approach that
we have followed in this review. It is only within a unified
theory encompassing general relativity and quantum mechanics that
such questions might be addressed. String theory may be such a
theory. The problem of the nature of the resolution of
cosmological singularities in string theory is still a vastly
unchartered  territory. There is a second kind of
singularity which arises when two branes collide. In such a case there is also a
singularity in the low energy effective action as one of the 
extra dimensions shrink to zero size. It was speculated that 
brane collisions play an important role in cosmology, especially in 
order to understand the big bang itself \cite{ekpyrotic1}-\cite{bornagain}.

In heteroric M-theory the regime where the distance between the branes 
becomes small corresponds to the regime where the string coupling constant 
becomes small and therefore a perturbative heterotic treatment may be available. In
particular for adiabatic processes the resulting small instanton
transition has been thoroughly studied (see e.g. \cite{transitionov} and 
\cite{transitionwi}). Here we would like to
present an analysis of such a collision and of the possible
outcome of such a collision.
A natural and intuitive phenomenon which may occur during  a
collision is the existence of a cosmological bounce.
Such objects are not avalaible in 4d under mild assumptions, and
therefore can be exhibited as a purely extra dimensional signature.

We will describe a $d$--dimensional theory with a scalar field
and gravity whose solutions present a cosmological singularity at
$t=0$. It turns out that this model is the low energy
approximation of a purely $(d+1)$ dimensional model where the
extra dimension is an interval with two boundary branes. The
singularity corresponds to the brane collision. In the $(d+1)$
dimensional picture, one can extend the motion of the branes
past each other, hence providing a continuation of the brane
motion after the collision. The $(d+1)$ dimensional space--time is
equivalent to an orbifold where the identification  between
space--time points is provided by a Lorentz boost. These spaces
are the simplest possible space--times with a singularity. As
with ordinary spatial orbifolds, one may try to define string
theory in such backgrounds and analyse the stringy resolution of
the singularity. Unfortunately, these orbifolds are not stable in
general relativity ruling them out as candidate stringy backgrounds.
Let us now briefly outline some of the arguments.  

We have already investigated the moduli space approximation for models
with a bulk scalar field. Here we will consider that at low energy the moduli space
consists of a single scalar field $\phi$ coupled to gravity
\begin{equation}
S=\int d^d x \sqrt {-g} \left(R-\frac{1}{2}(\partial \phi)^2\right).
\label{eff}
\end{equation}
Cosmological solutions with
\begin{equation}
ds^2= a^2(t)[-dt^2 + dx^idx_i]
\end{equation}
can be easily obtained
\begin{equation}
a=a(1)\vert t\vert ^{\frac{1}{d-2}},\ \phi=\phi (1)+ \epsilon
\sqrt{\frac{2(d-1)}{d-2}}\ln \vert t \vert, \label{sol}
\end{equation}
where $\epsilon=\pm 1$.
There are two branches corresponding to $t<0$ and $t>0$ 
connected by a singularity at $t=0$. 

So what is the extra dimensional origin of a such a model?
One can uplift the previous system to $(d+1)$ dimensions by defining
\begin{equation}
\psi= e^{\gamma \phi}
\end{equation}
and
\begin{equation}
\bar g_{\mu\nu}= \psi^{-4/(d-2)} g_{\mu\nu},
\end{equation}
where $\gamma=\sqrt{(d-2)/8(d-1)}$.
Consider now the purely gravitational $(d+1)$ dimensional theory with the
metric
\begin{equation}
ds^2_{d+1}=\psi^4dw^2 + \bar g_{\mu\nu}dx^\mu dx^\nu,
\end{equation}
where $w\in [0,1]$.
The two  boundaries  at $w=0$ and $w=1$  are 
boundary branes. The dimensional
reduction on the interval $w\in [0,1]$, i.e. integrating over the
extra dimension,  yields the effective
action  (\ref{eff})  provided one restricts the two fields $\psi(x^\mu)$ and $\bar
g_{\mu\nu}(x^\mu)$ to be dependent on $d$-dimensions only.

Let us now consider the nature of $(d+1)$ --dimensional
space-time obtained from the solutions (\ref{sol}).
The
$(d+1)$ dimensional metric becomes
\begin{equation}
ds_{d+1}^2= B^2 t^2 dw^2 + \eta_{\mu\nu}dx^\mu dx^\nu
\end{equation}
for a given $B$ depending on the integrations constants $\phi
(1)$ and $a(1)$. 
The geometry of space-time is remarkably simple.
It is a direct product $R^{d-1}\times M$ where $M$ is the
two dimensional compactified Milne space
whose metric is
\begin{equation}
ds_M^2= -dt^2+ B^4 t^2 dw^2.
\end{equation}
Using the light cone coordinates 
\begin{equation}
x^\pm= \pm t e^{\pm B^2w}
\end{equation}
the metric of Milne space reads
\begin{equation}
ds^2_M=dx_+dx_-,
\end{equation}
coinciding with the two dimensional Minkowski metric . There is
one subtlety here, the original identification  of the
extra--dimensional interval is here transcribed in the fact that
Milne space is modded out
by the boost
\begin{equation}
x^\pm \to e^{\pm 2  B^2}x^{\pm},
\end{equation}
as we have identified the interval with $S^1/Z_2$ and the
boundary branes are the fixed points of the $Z_2$ action as in
the Randall-Sundrum model.

The two boundary branes collide at $x^\pm =0$, their trajectories
are given by
\begin{equation}
x^{\pm}_0=\pm t, x_1^{\pm}= \pm t e^{\pm B^2}.
\end{equation}
At the singularity one can hope that the branes go past each other 
and evolve henceforth. Unfortunatly, Horowitz and Polchinski have 
shown that the structure of the orbifold space-time is unstable \cite{desaster}.
By considering a particle in this geometry, they showed that space--time
collapses to a space-like singularity.
Indeed one can focus on a particle and its $n$--th image under
the orbifold action. In terms of collision the impact parameter $b$
becomes constant as $n$ grows while the centre of mass energy
$\sqrt s$ grows like $\cosh nB^2$. As soon as $n$ is large
enough,
\begin{equation}
G \sqrt s > b^{d-2}
\end{equation}
the two particle approach each other within their Schwarzschild
radii therefore forming a black hole through gravitational collapse.
So the orbifold space--time does not make sense in general relativity,
i.e. not defining a time-dependent background for string theory.

Hence, it seems that the most simple example of brane collision
needs to be modified in order to provide a working example
of singularity with a meaningful string theoretic resolution.
It would be extremely relevant if one could find examples of
stable backgrounds of string theory where a cosmological
singularity can be resolved using stringy arguments. A
particularly promising avenue is provided by S--branes where a
cosmological singularity is shielded by a horizon 
(see e.g. \cite{quevedo} and references therein). 
Time will certainly tell
which of these approaches could lead to a proper understanding of cosmological
singularities and their resolutions, an issue highly
relevant to brane cosmology both in the early universe and the
recent past.

\section{Open Questions}
In this article we have reviewed different aspects of brane
cosmology in a hopefully pedagogical manner reflecting our own biased
point of view. Let us finally 
summarize some of the open questions.
\begin{itemize}
\item In the case of the single brane model by Randall \& Sundrum, 
the homogeneous cosmological evolution is well understood. An 
unsolved issue in this model, however, is a  complete understanding 
of the evolution of cosmological perturbations. The effects of the 
bulk gravitational field, encoded in the  projected Weyl--tensor, on  
CMB physics and Large Scale Structures are not known. The problem 
is twofold: first, the bulk equation are partial, non--linear differential 
equations and second, boundary conditions on the brane have to be imposed. 
The current formalism have not yet been used in order to tackle these problems
(for perturbations in brane world theories, see 
\cite{perturbationsfirst}-\cite{perturbationslast}; a short review on 
brane world perturbations is given in \cite{nathaliereview}).

\item Models with bulk scalar fields: Although we have presented some 
results on the cosmological evolution of a homogeneous brane, we assumed 
that the bulk scalar field does not strongly vary around the brane. Clearly, 
this needs to be investigated in some detail through a detailed investigation 
of the bulk equations, presumably with the help of numerical methods. Furthermore, 
for models with two branes, the cosmology has to be explored also in the high 
energy regime, in which the moduli--space approximation is not valid. Some exact 
cosmological solutions have been found in 
\cite{boundaryinflation} and \cite{skinner}.

\item Both the bulk scalar field as well as the interbrane distance in 
two brane models could play an 
important role at least during some part of the cosmological history. Maybe one 
of the fields plays the role of dark energy. In that case, it is only natural 
that masses of particles vary, as well as other parameter, such as the fine 
structure constant  $\alpha_{em}$ \cite{varyour}. Details of this interesting 
proposal have yet to be worked out.

\item The bulk singularity, which was thought to be shielded away with the help 
of a second brane, seems to play an important role in a cosmological setting. 
We have seen in section 5 that the negative tension brane moves towards 
the bulk singularity and eventually hits it. Therefore, cosmology forces us 
to think about this singularity, even if it was shielded with a second brane. 
Cosmologically, the brane might be 
repelled, which might be described by a potential. 
Alternatively, one might take quantum corrections into account in form of 
a Gauss-Bonnett term in the bulk \cite{gbfirst}-\cite{gblast}.

\item Brane collisions provide a different singularity problem in brane cosmology. 
String theory has to make progress in order to understand this singularity as well. 
From the cosmological point of view, the question is, if a transition between the 
brane collision can provide a new cosmological era and how cosmological perturbations 
evolve before and after the bounce \cite{bcpertfirst}-\cite{bcpertlast}. 
\end{itemize}

These aspects of brane cosmology raise very interesting questions.
It is clear that cosmology will continue to play an important role 
for testing our ideas beyond the standard model of particle physics. 

\noindent{\bf Acknowledgements:} We are grateful to our 
collaborators and colleagues P. Ashcroft, R. Battye, T. Boehm, R. Brandenberger,
P. Binetruy, C. Charmousis, A. Davis, S. Davis, M. Dorca, R. Durrer, C. Gordon, 
C. Grojean, R. Guedens, D. Langlois, A. Lukas, R. Maartens, C. Martins, A. Mennim, 
M. Parry, G. Palma, L. Pilo, C. Rhodes, M. Rodriguez-Martinez, J. Soda, D. Steer, 
N. Turok, D. Wands, S. Webster and T. Wiseman for many useful discussions and 
comments on brane cosmology. We are grateful to Anne Davis for her comments on an 
earlier draft of the paper. C.v.d.B. is supported by PPARC.


\section*{References}
\begin{thebibliography}{300}
\bibitem{Kaluza} Kaluza, T. 1921 {\it Sitzungsber.Preuss.Akad.
Wiss.Berlin (Math.Phys.) K1} p 966; Klein O 1926 Z.Phys. {\bf 37} 895
\bibitem{polchinskibook} Polchinski J. 1999 {\it String Theory, Two
Volumes}, Cambridge University Press
\bibitem{akama} Akama K 1982 Lect.Notes Phys {\bf 176} 267 
\bibitem{rubakov} Rubakov V A, Shaposhnikov M E 1983 Phys.Lett.B {\bf
125} 136
\bibitem{visser} Visser M 1985 Phys.Lett.B {\bf 159} 22
\bibitem{squires} Squires E J 1986 Phys.Lett.B {\bf 167} 286
\bibitem{gibbons} Gibbons G W, Wiltshire D L 1987 
Nucl. Phys. B{\bf 717} 
\bibitem{horavawitten} Horava P, Witten E 1996 Nucl.Phys.B{\bf 460} 
506; ibid Nucl.Phys.B{\bf 475} 94 
\bibitem{witten} Witten E 1996 Nucl. Phys.B {\bf 471} 135
\bibitem{lukasstelle} Lukas A, Ovrut B A, Stelle K S, Waldram D 1999 
Phys.Rev.D{\bf 59} 086001; ibid Nucl.Phys.B {\bf 552} 246
\bibitem{arkanihamed1} Arkani-Hamed N, Dimopoulos S, Dvali G 1998 
Phys.Lett.B{\bf 429} 263 
\bibitem{arkanihamed2} Antoniadis I, Arkani-Hamed N, Dimopoulos S, Dvali G 1998 
Phys.Lett.B{\bf 436} 257 
\bibitem{antoniadis} Antoniadis I 1990 Phys.Lett.B{\bf 246} 377
\bibitem{randallsundrum1} Randall L, Sundrum R 1999
Phys.Rev.Lett. {\bf 83} 3370 
\bibitem{randallsundrum2} Randall L, Sundrum R 1999
Phys.Rev.Lett. {\bf 83} 4690 
\bibitem{goldbergerwise} Goldberger W D, Wise M B 1999 Phys.Rev.Lett. 
{\bf 83} 4922
\bibitem{selftuning} Arkani-Hamed N, Dimopoulos S, Kaloper N, Sundrum R 
2000 Phys. Lett.B {\bf 480} 193
\bibitem{nilles} Forste S, Lalak Z, Lavignac S, Nilles H-P 2000 
Phys.Lett.B {\bf 481} 360
\bibitem{binetruy1} Binetruy P, Deffayet C, Langlois D 2000 
Nucl.Phys.{\bf 565} 269 
\bibitem{boundaryinflation} Lukas A, Ovrut B A, Waldram D 1999 
Phys.Rev.D {\bf 61} 
\bibitem{grojean} Cline J M, Grojean C, Servant G 1999 
Phys.Rev.Lett. {\bf 83} 4245 
\bibitem{csaki} Csaki C, Graesser M, Kolda C, Terning J 1999 
Phys.Lett.B {\bf 462} 34 
\bibitem{luki} Lukas A, Ovrut B A, Waldram D 1998 hep-th/9812052
\bibitem{rubi} Rubakov V A 2001 Phys.Usp.{\bf 44} 871
\bibitem{damien} Easson D A 2000 Int.J.Mod.Phys. A{\bf 16} 4803
\bibitem{wandi} Wands D 2002 Class.Quant.Grav.{\bf 19} 3403 
\bibitem{langi} Langlois D 2002 hep-th/0209261; to appear in the
proceedings of YITP Workshop: {\it Braneworld: Dynamics of Space-time 
Boundary}
\bibitem{quevedo} Quevedo F 2002 Class.Quant.Grav. {\bf 19} 5721
\bibitem{submillimeter} Hoyle C D, et al 2001 Phys.Rev.Lett. {\bf 86} 
1418 
\bibitem{flanagan} Flanagan E E, Tye S H H, Wasserman I 2000 
Phys.Rev.D {\bf 62} 044039 
\bibitem{vandebruck1} van de Bruck C, Dorca M, Martins C, Parry M 2000 
Phys.Lett.B {\bf 495} 183
\bibitem{darkradiation1} Mukohyama S 1999 Phys.Lett.B {\bf 473} 241
\bibitem{darkradiation2} Ida D 2000 JHEP {\bf 0009} 014
\bibitem{darkradiation3} Ichiki K, Yahiro M, Takino T, Orito M, 
Mathews G J 2002 Phys.Rev.D {\bf 66} 043521
\bibitem{shirubulk} Mukohyama S, Shiromizu T, Maeda K-I 2000 
Phys.Rev.D {\bf 62} 024028 
\bibitem{kraus} Kraus P 1999 JHEP {\bf 9912} 011
\bibitem{binetruy2} Binetruy P, Deffayett C, Ellwanger U, Langlois D 2000 
Phys.Lett.B {\bf 477} 285
\bibitem{charmousis} Bowcock P, Charmousis C, Gregory R 2000 
Class.Quant.Grav.{\bf 17} 4745
\bibitem{maartens} Maartens R 2001 gr-qc/0101059 
\bibitem{shiromizu} Shiromizu T, Madea K, Sasaki M 2000 
Phys.Rev.D {\bf 62} 024012
\bibitem{wald} Wald R 1984 {\it General Relativity} University of Chicago Press
\bibitem{israel} Israel W 1966 Nuovo Cim B{\bf 44S10} 1; 
Erratum: ibid Nuovo Cim B{\bf 48} 463 
\bibitem{roy} Maartens R 2000 Phys.Rev.D {\bf 62} 084023
\bibitem{wandsinflation} Maartens R, Wands D, Bassett B A, 
Heard I 2000 Phys.Rev.D{\bf 62} 041301
\bibitem{copeland} Copeland E J, Liddle A R, Lidsey J E 2001 
Phys.Rev.D {\bf 64} 023509
\bibitem{anne} Davis S C, Perkins W B, Davis A-C, Vernon I R 2000 
Phys.Rev.D{\bf 63} 083518
\bibitem{rsperturbations} Langlois D, Maartens R, Sasaki M, Wands D 2001 
Phys.Rev.D{\bf 63} 084009
\bibitem{bardeen} Bardeen J M, Steinhardt P J, Turner M S 
1983 Phys.Rev.D{\bf 28} 679
\bibitem{malik} Wands D, Malik K, Lyth D H, Liddle A R (2000) 
Phys.Rev.D{\bf 62} 043527 
\bibitem{liddlebook} Liddle A R, Lyth D 2000 {\it Cosmological 
Inflation and Large Scale Structure} Cambridge University Press 
\bibitem{langloisperturbations} Langlois D, Maartens R, Wands D 2000 
Phys.Lett.B {\bf 489} 259 
\bibitem{rubakovperturbations} Gorbunov D S, Rubakov V A, 
Sibiryakov S M (2001) J. High Energy Phys. JHEP 10(2001)015 
\bibitem{gordon} Gordon D, Wands D, Bassett B A, Maartens R (2001) 
Phys.Rev.D{\bf 63} 023506 
\bibitem{ashcroft} Ashcroft P, van de Bruck C, Davis A C 2002 
Phys.Rev.D {\bf 66} 121302
\bibitem{raf} Guedens R, Clancy D, Liddle A R 2002 
Phys.Rev.D{\bf 66} 043513; ibid 2002 Phys.Rev.D{\bf 66} 083509
\bibitem{gubser} Gubser S S 2001 Phys.Rev.D{\bf 63} 084017
\bibitem{csakiholo} Csaki C, Erlich J, Hollowood T J, Terning J 2001 
Phys.Rev.D{\bf 63} 065019
\bibitem{verlinde} Verlinde E 2000 hep-th/0008140
\bibitem{tanaka} Tanaka T 2002 gr-qc/0203082
\bibitem{kaloper1} Emparan R, Fabbri A, Kaloper N 2002 JHEP {\bf 0208} 043
\bibitem{kaloper2} Emparan R, Garcia-Bellido J, Kaloper N 2002 hep-th/0212132
\bibitem{hawking1} Hawking S W, Hertog T, Reall H S 2000 Phys.Rev.D{\bf 62} 043501
\bibitem{hawking2} Hawking S W, Hertog T, Reall H S 2001 Phys.Rev.D{\bf 63} 083504
\bibitem{tracefirst} Nojiri S, Odinstov S D 2000 Phys.Lett.B{\bf 484} 119
\bibitem{tracelast} Anchordoqui L, Nunez C, Olsen K 2000 JHEP {\bf 0010} 050
\bibitem{maedawands} Maeda K, Wands D 2000 Phys.Rev.D{\bf 62} 124009
\bibitem{mennim} Mennim A, Battye R A 2001 Class.Quant.Grav.{\bf 18} 2171
\bibitem{kanti1} Kanti P, Olive K A, Pospelov M 2000 
Phys.Lett.B{\bf 481} 386
\bibitem{langloisbulkscalar} Langlois D, Rodriguez-Martinez M 2001 
Phys.Rev. D{\bf 64} 123507
\bibitem{flanagan2} Flanagan E E, Tye S H H, Wasserman I 2001 
Phys.Lett.B{\bf 522} 155 
\bibitem{chamblin} Chamblin H A, Reall H S 1999 Nucl.Phys. {\bf 562} 133
\bibitem{kunze} Kunze K E, Vazquez-Mozo M A 2002 Phys.Rev.D{\bf 65} 044002 
\bibitem{stephen1} Davis S C 2002 JHEP {\bf 0203} 054
\bibitem{stephen2} Davis S C 2002 JHEP {\bf 0203} 058
\bibitem{kanti2} Kanti P, Lee S, Olive K A 2002 hep-th/0209036 
\bibitem{karch} DeWolfe O, Freedman D Z, Gubser S S, Karch A 2000 
Phys.Rev.D {\bf 62} 046008
\bibitem{brax1} Brax Ph, Davis A C 2001 Phys.Lett.B {\bf 497} 
289
\bibitem{groje} Csaki C, Erlich J, Grojean C, Hollowood T J 2000 
Nucl. Phys. B{\bf 584} 359 
\bibitem{youm} Youm D 2001 Nucl. Phys. B{\bf 596} 289
\bibitem{braxsingularity} Brax Ph, Davis A C 2001 Phys.Lett.B 
{\bf 513} 156
\bibitem{bscosfirst} Mohapatra R N, Perez-Lorenzana A, 
de Sousa Pires C A 2000 Phys.Rev.D{\bf 62} 105030
\bibitem{himo} Himemoto Y, Sasaki M 2002 
Phys.Rev.D {\bf 63} 044015
\bibitem{hime} Himemoto Y, Tanaka T, Sasaki M 2002 
Phys.Rev.D {\bf 65} 104020
\bibitem{koyama2} Koyama K, Takahashi K 2003 hep-th/0301165
\bibitem{bscoslast} Wang B, Xue L, Zhang X, Hwang W YP 2003 
hep-th/0301072
\bibitem{brax2} Brax Ph, Davis A C 2001 J. High Energy Phys. 
JHEP 05(2001)007 
\bibitem{vandebruck2} Brax Ph, van de Bruck C, Davis A C 2001 
J. High Energy Phys. JHEP 10(2001)026 
\bibitem{chiba} Chiba, T 2001 gr-qc/0110118 
\bibitem{uzan} Uzan J-P 2002 hep-ph/0205340, to appear in 
Rev. Mod. Physics 
\bibitem{fujiibook} Fujii Y, Maeda K-I 2003 {\it The Scalar-Tensor Theory 
of Gravitation} Cambridge University Press
\bibitem{kiritsis} Kiritsis E, Kofinas G, Tetradis N, Tomaras T N, Zarikas V 2002 
hep-th/0207060
\bibitem{tetradis} Tetradis N 2002 hep-th/0211200
\bibitem{braxlanglois} Brax Ph, Langlois D, Rodriguez-Martinez M 2002 
hep-th/0212067
\bibitem{garriga} Garriga J, Tanaka T 2000 Phys.Rev.Lett.{\bf 84} 2778 
\bibitem{chiba2} Chiba T 2000 Phys.Rev.D{\bf 62} 021502
\bibitem{brax3} Brax Ph, van de Bruck C, Davis A C, 
Rhodes C.S. 2002 Phys.Lett.B {\bf 531} 135
\bibitem{brax4} Brax Ph, van de Bruck C, Davis A C, 
Rhodes C.S. 2002 Phys.Rev.D {\bf 65} 121501 
\bibitem{vandebruck3} Brax Ph, van de Bruck C, Davis A C, 
Rhodes C.S. 2003 Phys.Rev.D{\bf 67} 023512
\bibitem{cynolter} Cynolter G 2002 hep-th/0209152
\bibitem{radcosfirst} Csaki C, Graesser M l, Randall L, Terning J 1999 
Phys.Rev.D {\bf 62} 045015
\bibitem{radioncharmousis} Charmousis C, Gregory R, Rubakov V A 2000 
Phys.Rev.D{\bf 62} 067505
\bibitem{greasser} Csaki C, Graesser M L, Kribs G D 2001 
Phys.Rev.D{\bf 63} 065002
\bibitem{gen} Gen U, Sasaki M 2001 Prog.Theor.Phys. {\bf 105} 591
\bibitem{nolte} Grinstein B, Nolte D R, Skiba W 2001 
Phys.Rev.D{\bf 63} 105016
\bibitem{langrad} Binetruy P, Deffayet C, Langlois D 2001 
Nucl.Phys.B{\bf 615} 219
\bibitem{anupam} Mazumdar A, Perez-Lorenzana A 2001 
Phys.Lett.B{\bf 508} 340 
\bibitem{radcoslast} Langlois D, Sorbo L 2002 Phys.Lett.B{\bf 543} 155
\bibitem{barrow} Webb J K, Murphy M T, Flambaum V V, Dzuba V A, 
Barrow J D, Churchill C W, Prochaska J X, Wolfe A M 2001 
Phys.Rev.Lett. {\bf 87} 091301
\bibitem{braneaction0} Khoury J, Zhang R 2002 
Phys.Rev.Lett.{\bf 89} 061302
\bibitem{braneaction1} Kanno S, Soda J 2002 
Phys.Rev.D{\bf 66} 083506
\bibitem{braneaction2} Kobayashi S, Koyama K 2002 JHEP {\bf 0212} 056 
\bibitem{braneaction3} Shiromizu T, Koyama K 2002 preprint 
[hep-th/0210066]
\bibitem{braneaction4} Wiseman T 2002 Class.Quant.Grav. {\bf 19} 
3083
\bibitem{gilles} Damour T, Esposito-Farese G 1992 
Class.Quant.Grav.{\bf 9} 2093
\bibitem{damour} Damour T, Nordtvedt K 1993 Phys.Rev.Lett.{\bf 70} 2217
\bibitem{ekpyrotic1} Khoury J, Ovrut B A, Steinhardt P J, Turok N 2001 
Phys.Rev.D{\bf 64} 123522
\bibitem{ekpyrotic2} Khoury J, Ovrut B A, Seiberg N, Steinhardt P J, Turok N 2002 
Phys.Rev.D{\bf 65} 086007
\bibitem{ekpyrotic3} Khoury J, Ovrut B A, Steinhardt P J, Turok N 2002 
Phys.Rev.D{\bf 66} 046005 
\bibitem{ekpyrotic4} Enqvist K, Keski-Vakkuri E, Rasanen S 2001 
Phys.Rev.D {\bf 614} 388
\bibitem{ekpyrotic5} Kallosh R, Kofman L, Linde A D 2001 
Phys.Rev.D {\bf 64} 123523
\bibitem{nero} Neronov A 2001 JHEP {\bf 0111} 007
\bibitem{rasanen} Rasanen S 2002 Nucl.Phys. B{\bf 626} 183
\bibitem{cyclic} Steinhardt P J, Turok N 2002 
Science {\bf 296} 1436; ibid 2002 Phys.Rev.D {\bf 65} 126003
\bibitem{bornagain} Kanno S, Sasaki M, Soda J 2002 preprint hep-th/0210250 
\bibitem{transitionov} Ovrut B A, Pantev T, Park J 2000 JHEP {\bf 0005} 045
\bibitem{transitionwi} Witten E 1996 Nucl. Phys. B{\bf 460} 541
\bibitem{desaster} Horowitz G T, Polchinski J 2002 Phys.Rev.D{\bf 66} 103512
\bibitem{perturbationsfirst} Mukohymama S 2000 Phys.Rev.D {\bf 62} 084015
\bibitem{kodama} Kodama H, Ishibashi A, Seto O 2000 Phys.Rev.D {\bf 62} 064022
\bibitem{roypertur} Maartens R 2000 Phys.Rev.D {\bf 62} 084023
\bibitem{langlispertur} Langlois D 2000 Phys.Rev.D {\bf 62} 126012
\bibitem{vdbruckpertur} van de Bruck C, Dorca M, Brandenberger R H, Lukas A 2000 
Phys.Rev.D {\bf 62} 123515
\bibitem{koyama} Koyama K, Soda J 2000 Phys.Rev.D {\bf 62} 123502
\bibitem{royandchris} Gordon C, Maartens R 2001 Phys.Rev.D{\bf 63} 044022
\bibitem{langloisnew} Langlois D 2001 Phys.Rev.Lett. {\bf 86} 2212
\bibitem{malik2} Bridgman H A, Malik K A, Wands D 2002 Phys.Rev.D {\bf 65} 043502
\bibitem{nathalie} Deruelle N, Dolezel T, Katz J 2001 Phys.Rev.D {\bf 63} 083513
\bibitem{vdb2} van de Bruck C, Dorca M 2000 hep-th/0012073 
\bibitem{vdb3} Dorca M, van de Bruck C 2001 Nucl.Phys.B {\bf 605} 215
\bibitem{malik3} Bridgman H A, Malik K A, Wands D 2001 Phys.Rev.D {\bf 63} 084012
\bibitem{soda2} Koyama K, Soda J 2002 Phys.Rev.D{\bf 65} 023514
\bibitem{bernhard1} Leong B, Dunsby P, Challinor A, Lasenby A 2002 
Phys.Rev.D{\bf 65} 104012
\bibitem{bernhard2} Leong B, Challinor A, Maartens R, Lasenby A 2002 
Phys.Rev.D{\bf 66} 104010
\bibitem{deffayet} Deffayet C 2002 Phys.Rev.D {\bf 66} 103504 
\bibitem{riotti} Giudice G F, Kolb E W, Lesgourges J, Riotto A 2002 
Phys.Rev.D {\bf 66} 083512
\bibitem{perturbationslast} Riazuelo A, Vernizzi F, Steer D, Durrer R 2002 
hep-th/0205220
\bibitem{nathaliereview} Deruelle N 2003 gr-qc/0301035
\bibitem{skinner} Lukas A, Skinner D 2001 JHEP {\bf 0109} 020
\bibitem{varyour} Brax Ph, van de Bruck C, Davis A-C, Rhodes C S 2003 
hep-th/0210057; to appear in Astrophys. and Space Science
\bibitem{gbfirst} Binetruy P, Charmousis C, Davis S C, Dufaux J-F 2002 
Phys.Lett.B {\bf 544} 183 
\bibitem{germani} Germani C, Sopuerta C F 2002 Phys.Rev.Lett. {\bf 88} 231101
\bibitem{gbchristos} Charmousis C, Dufaux J-F 2002 Class.Quant.Grav.{\bf 19} 4671 
\bibitem{gblast} Gravanis E, Willison S 2002 hep-th/0209076
\bibitem{bcpertfirst} Durrer R, Vernizzi F 2002 Phys.Rev.D {\bf 66} 083503
\bibitem{peter} Peter P, Pinto-Neto N 2002 Phys.Rev.D {\bf 66} 063509 
\bibitem{brand} Brandenberger R, Finelli F 2001 JHEP{\bf 11} 056
\bibitem{lyth} Lyth D H 2002 Phys.Lett.B{\bf 524} 1
\bibitem{hwang} Hwang J-C 2002 Phys.Rev.D {\bf 65} 063514
\bibitem{branden} Tsujikawa S, Brandenberger R H, Finelli F 2002 
Phys.Rev.D {\bf 66} 083513
\bibitem{bcpertlast} Cartier C, Durrer R, Copeland E J 2003 hep-th/0301198
\end{thebibliography}
\end{document}